\documentclass[12pt]{article}
\usepackage{amssymb}
\usepackage{amsmath}
\usepackage{graphicx,color}
\usepackage{cite}
\usepackage[subrefformat=parens]{subcaption}

\numberwithin{equation}{section}

\begin{document}
\begin{titlepage}
%
\begin{center}
\baselineskip 25pt 
{\Large\bf
Right-handed neutrino dark matter in the classically conformal U(1)$^\prime$ extended Standard Model
}
\end{center}
\vspace{2mm}
\begin{center}
{\large
Satsuki Oda$^{~a,b}$ \footnote{satsuki.oda@oist.jp},
Nobuchika Okada$^{~c}$ \footnote{okadan@ua.edu}, 
and 
Dai-suke Takahashi$^{~a,b}$ \footnote{daisuke.takahashi@oist.jp}
}
\end{center}
\vspace{2mm}

\begin{center}
{\it
$^{a}$Okinawa Institute of Science and Technology Graduate University (OIST), \\ 
Onna, Okinawa 904-0495, Japan
\vspace{5mm}

$^{b}$Research Institute, Meio University, \\
Nago, Okinawa 905-8585, Japan \\
\vspace{5mm}

$^{c}$Department of Physics and Astronomy, University of Alabama, \\
Tuscaloosa, Alabama 35487, USA
}
\end{center}
\vspace{3mm}
\begin{abstract}
We consider the dark matter (DM) scenario in the context
   of the classically conformal U(1)$^\prime$ extended standard model (SM),
   with three right-handed neutrinos (RHNs) and the U(1)$^\prime$ Higgs field.
The model is free from all the U(1)$^\prime$ gauge and gravitational anomalies
   in the presence of the three RHNs.
We introduce a $Z_2$-parity in the model,
   under which an odd-parity is assigned to one RHN,
   while all the other particles is assigned to be $Z_2$-even,
   and hence the $Z_2$-odd RHN serves as a DM candidate.
In this model, the U(1)$^\prime$ gauge symmetry is radiatively broken
   through the Coleman-Weinberg mechanism,
   by which the electroweak symmetry breaking is triggered.
There are three free parameters in our model,
   the U(1)$^\prime$ charge of the SM Higgs doublet ($x_H$),
   the new U(1)$^\prime$ gauge coupling ($g_X$),
   and the U(1)$^\prime$ gauge boson ($Z^\prime$) mass ($m_{Z^\prime}$),
   which are severely constrained
   in order to solve the electroweak vacuum instability problem,
   and satisfy the LHC Run-2 bounds from the search for $Z^\prime$ boson resonance.
In addition to these constraints, we investigate the RHN DM physics.
Because of the nature of classical conformality, 
   we find that a RHN DM pair mainly annihilates into the SM particles
   through the $Z^\prime$ boson exchange.
This is the so-called $Z^\prime$-portal DM scenario.
Combining the electroweak vacuum stability condition, the LHC Run-2 bounds,
   and the cosmological constraint from the observed DM relic density,
   we find that all constrains complementarily work to narrow down the allowed parameter regions,
   and, especially, exclude $m_{Z^\prime} \lesssim 3.5$ TeV.
For the obtained allowed regions,
   we calculate the spin-independent cross section of the RHN DM with nucleons.
We find that the resultant cross section well below the current experimental upper bounds.

\end{abstract}
\end{titlepage}

\section{Introduction}
There are important missing pieces in the Standard Model (SM),
   for example, a candidate for the dark matter (DM), and tiny neutrino masses and their flavor mixings. 
The SM should be extended so as to supplement these missing pieces. 
The so-called seesaw mechanism is a natural way
   to reproduce the tiny neutrino masses~\cite{Minkowski:1977sc,Yanagida:1979as,
   GellMann:1980vs,Glashow:1979nm,Mohapatra:1979ia},
   where heavy Majorana right-handed neutrinos (RHNs) are introduced.
The minimal gauged $B-L$ model~\cite{Mohapatra:1980qe,Marshak:1979fm,Wetterich:1981bx,
   Masiero:1982fi,Mohapatra:1982xz,Buchmuller:1991ce}
   is one of the simplest extensions of the SM with an extra gauge symmetry,
   in which the accidentally anomaly-free global $B-L$ (baryon number minus lepton number) in the SM
   is gauged.
Three RHNs play an essential roll to cancel the gauge and gravitational anomalies of the model.
Associated with the $B-L$ symmetry breaking, the RHNs acquire their Majorana masses,
   and hence the seesaw mechanism is automatically implemented.
The minimal $B-L$ model can be generalized
   to the so-called minimal U(1)$^\prime$ model~\cite{Appelquist:2002mw}.
Here, the U(1)$^\prime$ gauge group is defined as a linear combination of the U(1)$_{B-L}$
   and the SM U(1)${_Y}$ gauge groups,
   so that the U(1)$^\prime$ model is anomaly-free.

In our previous work~\cite{Oda:2015gna,Das:2016zue},
   we have investigated the minimal U(1)$^\prime$ model with classically conformal invariance.\footnote{
See Refs.~\cite{Hempfling:1996ht,Dias:2006th,Espinosa:2007qk,Chang:2007ki,Foot:2007as,
   Foot:2007ay,Meissner:2006zh,Foot:2007iy,Meissner:2007xv,Meissner:2008gj,Iso:2009ss,
   Iso:2009nw,Holthausen:2009uc,Farzinnia:2013pga,Heikinheimo:2013fta,Farzinnia:2014xia,
   Lindner:2014oea,Khoze:2014xha,Gabrielli:2013hma,Altmannshofer:2014vra,Karam:2015jta,
   Haba:2015yfa,Okada:2015gia,Latosinski:2015pba,Wang:2015sxe,Goertz:2015dba,Haba:2015lka,
   Haba:2015nwl,Ghorbani:2015xvz,Haba:2015qbz,Ahriche:2015loa,Ishida:2016ogu,Hatanaka:2016rek,
   Karam:2016rsz,Marzola:2016xgb,Das:2015nwk,Kannike:2016wuy,Marzola:2017jzl}
   for recent work on new physics models with classically conformal invariance.}
In this model, the U(1)$^\prime$ gauge symmetry is radiatively broken
   through the Coleman-Weinberg (CW) mechanism~\cite{Coleman:1973jx}.
Given a negative mixing quartic coupling between the SM Higgs and the U(1)$^\prime$ Higgs fields, 
   once the U(1)$^\prime$ Higgs field develops a vacuum expectation value (VEV),
   a negative mass squared of the SM Higgs doublet is generated,
   and thus the electroweak symmetry breaking is naturally triggered.
In this model context, we have investigated the electroweak vacuum instability problem in the SM. 
Employing the renormalization group (RG) equations at the two-loop level 
   and the central values for the world average masses
   of the top quark ($m_t=173.34$ GeV~\cite{ATLAS:2014wva}) 
   and the Higgs boson ($m_h=125.09$ GeV~\cite{Aad:2015zhl}),
   we have performed parameter scans to identify 
   the parameter region for resolving the electroweak vacuum instability problem. 
We have also investigated the ATLAS and CMS search limits at the LHC Run-2 (2015)  
   for the U(1)$^\prime$ gauge boson ($Z^\prime$)~\cite{TheATLAScollaboration:2015jgi,CMS:2015nhc},
   and identified the allowed parameter regions in our model. 
Combining the constraints from the electroweak vacuum stability and the LHC Run-2 results, 
   we have found a lower bound on the $Z^\prime$ boson mass.   
We also have calculated self-energy corrections to the SM Higgs doublet field through the heavy states, 
   the right-handed neutrinos and the $Z^\prime$ boson, and have found the naturalness bound 
   as $m_{Z^\prime} \lesssim 6$ TeV, in order to reproduce the right electroweak scale 
   for the fine-tuning level better than 10\%.

The so-called weakly interacting massive particle (WIMP) is one of the most promising candidates
   of the DM in our Universe,
   which is in thermal equilibrium in the early Universe. 
Among many possibilities, a simple way to introduce a WIMP DM
   in the minimal U(1)$^\prime$ model has been proposed
   in~\cite{Okada:2010wd} (see also \cite{Anisimov:2008gg}),
   where $Z_2$-parity is introduced and an odd-parity is assigned to one RHN,
   while all the other particles is assigned to be $Z_2$-even. 
We adapt this scheme in our minimal U(1)$^\prime$ model with the classically conformal invariance,
   and the $Z_2$-odd RHN is a DM candidate, 
   while the other two RHNs are utilized for the seesaw mechanism.
Note that only two RHNs are sufficient to reproduce the neutrino oscillation data,
   and the observed baryon asymmetry in the Universe
   through leptogenesis~\cite{Fukugita:1986hr}. 
This system is called the minimal seesaw \cite{King:1999mb,Frampton:2002qc}. 
In our model, there are two ways for the RHN DM to interact with the SM particles. 
One is mediated by the $Z^\prime$ boson ($Z^\prime$-portal)
   and the other is by the two Higgs bosons (Higgs portal)
   which are two mass eigenstates consisting of the SM Higgs and the U(1)$^\prime$ Higgs bosons. 
Recently, the $Z^\prime$-portal DM scenarios~\cite{Burell:2011wh,Basso:2012ti,Dudas:2013sia,Das:2013jca,Chu:2013jja,
   Lindner:2013awa,Alves:2013tqa,Kopp:2014tsa,Agrawal:2014ufa,Hooper:2014fda,
   Ma:2014qra,Alves:2015pea,Ghorbani:2015baa,Sanchez-Vega:2015qva,Duerr:2015wfa,
   Alves:2015mua,Ma:2015mjd,Okada:2016gsh,Okada:2016tzi,Chao:2016avy,Biswas:2016ewm,
   Accomando:2016sge,Fairbairn:2016iuf,Klasen:2016qux,Dev:2016xcp,Altmannshofer:2016jzy,
   Okada:2016tci,Kaneta:2016vkq,Arcadi:2017kky}
   have been intensively investigated,
   while the Higgs portal RHN DM scenarios~\cite{Okada:2010wd,Okada:2012sg,Basak:2013cga}
   have been analyzed in detail.

In this paper, we consider the classically conformal U(1)$^\prime$ extended SM with the RHN DM.
As we mentioned above, the allowed parameter regions in the classically conformal model
   are severely constrained
   in order to solve the electroweak vacuum instability problem,
   and to satisfy the LHC limits from the search for $Z^\prime$ boson resonance.
In addition to these constraints, we will investigate the RHN DM physics.
Because of the nature of classical conformality, 
   we find the mass mixing between the SM Higgs and the U(1)$^\prime$ Higgs bosons is very small,
   so that the RHN DM pair annihilation process mediated by the Higgs bosons is highly suppressed. 
Therefore, we focus on the study of the $Z^\prime$-portal RHN DM~\cite{Okada:2016gsh,Okada:2016tci}, 
   and identify allowed parameter regions to reproduce the observed DM relic density   
   from the Planck 2015 result~\cite{Aghanim:2015xee}.
We will show that the DM physics, LHC phenomenology, 
   and the electroweak vacuum stability condition complementarily work to narrow down
   the allowed parameter regions.
For the identified allowed regions,
   we also calculate the spin-independent cross section of the RHN DM with nucleons
   and compare our results with the current upper bounds from the direct DM search experiments.

This paper is organized as follows: 
In the next section, we introduce the classically conformal U(1)$^\prime$ extended SM
   with $Z^\prime$-portal RHN DM. 
We briefly review our previous work
   on the classically conformal U(1)$^\prime$ model~\cite{Oda:2015gna, Das:2016zue}.
In Sec.~\ref{Sec_relic_density}, we calculate the relic density of the $Z^\prime$-portal RHN DM. 
In Sec.~\ref{Sec_collider}, we study the $Z^\prime$ boson production at the LHC Run-2 (2016)~\cite{ATLAS:2016cyf,CMS:2016abv}, 
   and obtained the constraints on the model parameter space from the search results of the $Z^\prime$ boson resonance by the ATLAS and the CMS Collaborations. 
In Sec.~\ref{Sec_allowed_region}, we combine all the results in the previous sections
   and narrow allowed regions.
In Sec.~\ref{Sec_direct_detection}, for the allowed parameter regions,
   we calculate the spin-independent cross section of the RHN DM with nucleons.
The last section is devoted to conclusions.

\section{The classically conformal U(1)$^{\prime}$ extended SM with RHN DM}
\label{Sec_U1p}
In this section, we will briefly review the results in Ref.~\cite{Das:2016zue}.
Although the model is extended to incorporate the RHN DM, 
   the results presented here are essentially the same as those in Ref.~\cite{Das:2016zue}.

\subsection{The model}

\begin{table}[t]
\begin{center}
\begin{tabular}{c|ccc|rcr|c}
            & SU(3)$_c$ & SU(2)$_L$ & U(1)$_Y$ & \multicolumn{3}{c|}{U(1)$^\prime$} & $Z_2$ \\
\hline
&&&&&&&\\[-12pt]
$q_L^i$    & {\bf 3}   & {\bf 2}& $+1/6$ & $x_q$ 	& = & $\frac{1}{3}x_H + \frac{1}{6}x_\Phi$ &+ \\[2pt] 
$u_R^i$    & {\bf 3} & {\bf 1}& $+2/3$ & $x_u$ 		& = & $\frac{4}{3}x_H + \frac{1}{6}x_\Phi$  &+ \\[2pt] 
$d_R^i$    & {\bf 3} & {\bf 1}& $-1/3$ & $x_d$ 		& = & $-\frac{2}{3}x_H + \frac{1}{6}x_\Phi$  &+ \\[2pt] 
\hline
&&&&&&&\\[-12pt]
$\ell_L^i$    & {\bf 1} & {\bf 2}& $-1/2$ & $x_\ell$ 	& = & $- x_H - \frac{1}{2}x_\Phi$  &+  \\[2pt] 
$\nu_R^{1,2}$   & {\bf 1} & {\bf 1}& $0$   & $x_\nu$ 	& = & $- \frac{1}{2}x_\Phi$  &+\\[2pt] 
$\nu_R^3$   & {\bf 1} & {\bf 1}& $0$   & $x_\nu$ 	& = & $- \frac{1}{2}x_\Phi$  & $-$ \\[2pt] 
$e_R^i$   & {\bf 1} & {\bf 1}& $-1$   & $x_e$ 		& = & $- 2x_H - \frac{1}{2}x_\Phi$  &+ \\[2pt] 
\hline
&&&&&&&\\[-12pt]
$H$         & {\bf 1} & {\bf 2}& $+1/2$  &  $x_H$ 	& = & $x_H$\hspace*{12.5mm}  &+ \\ 
$\Phi$      & {\bf 1} & {\bf 1}& $0$  &  $x_\Phi$ 	& = & $x_\Phi$  &+ \\ 
\end{tabular}
\end{center}
\caption{
Particle contents of the U(1)$^\prime$ extended SM with $Z_2$ parity.
In addition to the SM particle contents, three generations of RHNs $\nu_R^i$
   ($i=1,2,3$ denotes the generation index) and U(1)$^\prime$ Higgs field $\Phi$ are introduced.
Under $Z_2$ parity, the only one RHN $\nu_R^3$ is odd,
  while the other particles, including $\nu_R^1$ and $\nu_R^2$, are even.
}
\label{Tab:particle_contents}
\end{table}

The model we will investigate is the anomaly-free U(1)$^\prime$ extension of the SM 
   with the classically conformal invariance, which is based on the gauge group 
   SU(3)$_C \times$SU(2)$_L \times$U(1)$_Y \times$U(1)$^\prime$. 
The particle contents of the model are listed in Table~\ref{Tab:particle_contents}. 
In addition to the SM particle content, three generations of RHNs $\nu_R^i$ 
   and a U(1)$^\prime$ Higgs field $\Phi$ are introduced.
We also introduce the $Z_2$ parity~\cite{Okada:2010wd},
   and assign an odd parity to one RHN $\nu_R^3$,
   while the other particles, including $\nu_R^1$ and $\nu_R^2$, have even parity. 
The conservation of $Z_2$ parity ensures the stability of $\nu_R^3$, which is 
   a unique candidate of the DM in our model.

The covariant derivative, which is relevant to U(1)$_Y \times$ U(1)$^\prime$, is defined as 
\begin{equation}
D_\mu \equiv \partial_\mu  
			- i\left(\begin{array}{cc} Y_{1} & Y_{X} \end{array}\right )
			\left(\begin{array}{cc} g_{1} & g_{1X} \\g_{X1} & g_{X} \end{array}\right)
			\left(\begin{array}{c} B_{\mu} \\B_{\mu}^\prime \end{array}\right), 
 \label{Eq:covariant_derivative}
\end{equation}
  where $Y_{1}$ ($Y_{X}$) is U(1)$_Y$ (U(1)$^\prime$ ) charge of a particle, 
  and the gauge coupling $g_{X1}$ and $g_{1X}$ are introduced associated with a kinetic mixing 
  between the two U(1) gauge bosons. 
In order to reproduce observed fermion masses  and flavor mixings, 
  we introduce the following Yukawa interactions: 
\begin{eqnarray}
{\cal L}_{\rm Yukawa} 
	&=& - \sum_{i=1}^{3} \sum_{j=1}^{3} Y_u^{ij} \overline{q_L^i} \tilde{H} u_R^j
		- \sum_{i=1}^{3} \sum_{j=1}^{3} Y_d^{ij} \overline{q_L^i} H d_R^j 
		- \sum_{i=1}^{3} \sum_{j=1}^{3} Y_e^{ij} \overline{\ell_L^i} H e_R^j \nonumber \\
	&& - \sum_{i=1}^{3} \sum_{j=1}^{2} Y_\nu^{ij} \overline{\ell_L^i} \tilde{H} \nu_R^j 
				- \sum_{i=1}^{3} Y_M^i \Phi \overline{\nu_R^{ic}} \nu_R^i + {\rm h.c.},
\label{Eq:L_Yukawa}
\end{eqnarray}
where $\tilde{H} \equiv i  \tau^2 H^*$, and the fourth and fifth terms in the right-handed side 
  are for the seesaw mechanism to generate neutrino masses. 
Without loss of generality, the Majorana Yukawa couplings in the fifth term
  are already diagonalized in our basis.
Because of the $Z_2$ parity, only two generation RHNs are involved 
  in the neutrino Dirac Yukawa couplings
  and hence the neutrino Dirac mass matrix is 2 by 3.
Once the U(1)$^\prime$ Higgs field $\Phi$ develops a VEV,
  the U(1)$^\prime$ symmetry is broken
  and the Majorana mass terms for the RHNs are generated.
After the electroweak symmetry breaking,
  the seesaw mechanism~\cite{Minkowski:1977sc,Yanagida:1979as,
  GellMann:1980vs,Glashow:1979nm,Mohapatra:1979ia} is automatically implemented,
  except that only two generation RHNs are relevant.
This system is the minimal seesaw~\cite{King:1999mb, Frampton:2002qc},
   which possesses a number of free parameters
  $Y_{\nu}^{ij}$ and $Y_M^{j}$ ($i=1,2,3$, $j=1,2$) enough
  to reproduce the neutrino oscillation data with a prediction of one massless eigenstate.

In the particle contents, the two parameters ($x_H$ and $x_\Phi$) reflect the fact that the U(1)$^\prime$  gauge group can be defined
  as a linear combination 
  of the SM U(1)$_Y$ and the U(1)$_{B-L}$ gauge groups. 
Since the U(1)$^\prime$ gauge coupling $g_{X}$ is a free parameter of the model and 
  it always appears as a product $x_\Phi g_{X}$ or  $x_H g_{X}$,
  we fix $x_\Phi=2$ without loss of generality throughout this paper. 
This convention excludes the case that U(1)$^\prime$ gauge group is identical with the SM U(1)$_Y$. 
The choice of $(x_H, x_\Phi)=(0, 2)$ corresponds to the U(1)$_{B-L}$ model. 
Another example is $(x_H, x_\Phi)=(-1, 2)$, which corresponds to the SM with the so-called U(1)$_R$ symmetry. 
When we choose $(x_H, x_\Phi)=(-16/41, 2)$, the beta function of $g_{X1}$ ($g_{1X}$) at the 1-loop level 
   has only terms proportional to $g_{X1}$ ($g_{1X}$)~\cite{Oda:2015gna}. 
This is the orthogonal condition between the U(1)$_Y$ and U(1)$^\prime$ at the 1-loop level, 
   under which  $g_{X1}$ and $g_{1X}$ do not evolve once we have set  $g_{X1}=g_{1X}=0$ at an energy scale.

Imposing the classically conformal invariance, the scalar potential is given by
\begin{equation}
V = \lambda_H \! \left( H^\dagger H \right)^2 
	+ \lambda_\Phi \! \left( \Phi^\dagger \Phi \right)^2 
	+ \lambda_{\rm mix} \! \left( H^\dagger H \right) \! \left( \Phi^\dagger \Phi \right) , 
\label{Eq:classical_potential}
\end{equation}
where the mass terms are forbidden by the conformal invariance. 
If $\lambda_{\rm mix}$ is negligibly small, 
  we can analyze the Higgs potential separately for $\Phi$ and $H$ as a good approximation. 
This will be justified in the following subsections. 
When the Majorana Yukawa couplings $Y_M^i$ are negligible compared to the U(1)$^\prime$ gauge coupling, 
 the $\Phi$ sector is identical with the original CW model \cite{Coleman:1973jx}, 
   so that the radiative U(1)$^\prime$ symmetry breaking will be achieved. 
Once $\Phi$ develops a VEV through the CW mechanism, 
   the tree-level mass term for the SM Higgs doublet is effectively generated through $\lambda_{\rm mix}$ 
   in Eq.~(\ref{Eq:classical_potential}).
Taking $\lambda_{\rm mix}$ negative, the induced mass squared for the Higgs doublet 
  is negative and, as a result, the electroweak symmetry breaking is driven in the same way as in the SM.

\subsection{Radiative U(1)$^\prime$ gauge symmetry breaking}

Assuming $\lambda_{\rm mix}$ is negligibly small, we first analyze the U(1)$^\prime$ Higgs sector. 
Without mass terms, the Coleman-Weinbeg potential \cite{Coleman:1973jx}
   at the 1-loop level is found to be 
\begin{eqnarray}
  V(\phi) =  \frac{\lambda_\Phi}{4} \phi^4 
 + \frac{\beta_\Phi}{8} \phi^4 \left(  \ln \left[ \frac{\phi^2}{v_\phi^2} \right] - \frac{25}{6} \right), 
\label{Eq:CW_potential} 
\end{eqnarray}
where $\phi / \sqrt{2} = \Re[\Phi]$, and 
  we have chosen the renormalization scale to be the VEV  
  of $\Phi$ ($\langle \phi \rangle =v_\phi$).  
Here, the coefficient of the 1-loop quantum corrections is given by 
\begin{eqnarray}
\beta_\Phi	&=& \frac{1}{16 \pi^2} 
		\left[ 20\lambda_\Phi^2 
			+ 6 x_\Phi^4 \left ( g_{X1}^2 + g_{X}^2 \right)^2 - 16\sum_i(Y_M^i)^4 \right]  \nonumber \\
	& \simeq &  \frac{1}{16 \pi^2} 
		\left[ 6 \left(x_\Phi g_{X} \right)^4 - 16\sum_i(Y_M^i)^4 \right] , 
\end{eqnarray}
  where in the last expression, we have used 
  $\lambda_\Phi^2 \ll (x_\Phi g_{X})^4$  as usual in the CW mechanism 
   and set $g_{X1}=g_{1X}= 0$ at $ \langle \phi \rangle = v_\phi$,  for simplicity. 
The stationary condition $\left. dV/d\phi\right|_{\phi=v_\phi} = 0$ leads to 
\begin{eqnarray}
   \lambda_\Phi = \frac{11}{6} \beta_\Phi, 
\label{eq:stationary}
\end{eqnarray} 
 and this $\lambda_\Phi$ is nothing but a renormalized self-coupling at $v_\phi$ defined as 
\begin{eqnarray}
 \lambda_\Phi = \frac{1}{3 !}\left. \frac{d^4V(\phi)}{d \phi^4} \right|_{\phi=v_\phi}. 
\end{eqnarray}  
For more detailed discussion, see Ref.~\cite{Khoze:2014xha}.

Associated with this radiative U(1)$^\prime$ symmetry breaking 
   (as well as the electroweak symmetry breaking),  
  the U(1)$^\prime$ gauge boson ($Z^\prime$ boson), 
  the Majorana RHNs $\nu_R^{1,2}$,
  and the RHN DM particle $\nu_R^3$ acquire their masses as 
\begin{eqnarray}
  m_{Z^\prime}  = \sqrt{(x_\Phi g_{X} v_\phi)^2  +  (x_H g_{X} v_h)^2} \simeq   x_\Phi g_{X} v_\phi, 
   \;  \;  m_{N^{1,2}} = \sqrt{2} Y_M^{1,2} v_\phi, 
   \;  \;  m_{\rm DM} = \sqrt{2} Y_M^3 v_\phi, 
\label{Eq:mass_Zp_DM}
\end{eqnarray} 
where $v_h=246$ GeV is the SM Higgs VEV, and we have used $x_\Phi v_\phi  \gg x_H v_h$, 
   which will be verified below. 
In this paper, we assume degenerate masses for $\nu_R^{1,2}$,  
   ($Y_M^1 = Y_M^2 = y_M$, equivalently, $m_{N^{1,2}}=m_N$), for simplicity. 
The U(1)$^\prime$ Higgs boson mass is given by 
\begin{eqnarray}
  m_\phi^2 &=& \left. \frac{d^2 V}{d\phi^2}\right|_{\phi=v_\phi}  
                    =\beta_\Phi v_\phi^2  \simeq 
  \frac{1}{8 \pi^2} \left( 3(x_\Phi g_{X})^4 - 16 y_M^4  - 8 y_{\rm DM}^4\right) v_\phi^2 \nonumber \\
  &\simeq&  \frac{1}{8 \pi^2}  \frac{ 3m_{Z^\prime}^4 - 4 m_N^4 - 2 m_{\rm DM}^4}{v_\phi^2},
\label{Eq:mass_phi}
\end{eqnarray} 
   where $y_{\rm DM} = Y_M^3$.
When the Yukawa couplings are negligibly small, this equation reduces to the well-known relation 
  derived in the original paper by Coleman-Weinberg \cite{Coleman:1973jx}. 
For a sizable Majorana mass, this formula indicates that 
  the potential minimum disappears, 
  so that there is an upper bound on the RHN mass 
  for the U(1)$^\prime$ symmetry to be broken radiatively.  
This is in fact the same reason why the CW mechanism
   in the SM Higgs sector fails to break the electroweak symmetry 
   when the top Yukawa coupling is large as observed. 
In order to avoid the destabilization of the U(1)$^\prime$ Higgs potential, 
   we simply set $m_{Z^\prime}^4 \gg m_N^4$ in the following analysis,
   while $m_{\rm DM} \simeq m_{Z^\prime}/2$ as we will find in the next section.
Note that this condition does not mean that the Majorana RHNs must be very light, 
   even though a factor difference between $m_{Z^\prime}$ and $m_N$ is enough to satisfy the condition. 
For simplicity, we set $y_M=0$ at $v_\phi$ in the following RG analysis as an approximation.

\subsection{Electroweak symmetry breaking}

Let us now consider the SM Higgs sector. 
In our model, the electroweak symmetry breaking is achieved in a very simple way. 
Once the U(1)$^\prime$ symmetry is radiatively broken, 
  the SM Higgs doublet mass is generated through the mixing quartic term between $H$ and $\Phi$ 
  in the scalar potential in Eq.~(\ref{Eq:classical_potential}), 
\begin{equation}
  V(h) = \frac{\lambda_H}{4}h^4 + \frac{\lambda_{\rm mix}}{4} v_\phi^2 h^2,  
\end{equation}
where we have replaced $H$ by $H = 1/\sqrt{2}\, (0,\,h)$ in the unitary gauge. 
Choosing $\lambda_{\rm mix} < 0$, the electroweak symmetry is broken in the same way
   as in the SM \cite{Iso:2009ss, Iso:2009nw}. 
However, we should note that a crucial difference from the SM is that, in our model, 
  the electroweak symmetry breaking originates from the radiative breaking
  of the U(1)$^\prime$ gauge symmetry. 
At the tree level, the stationary condition $V^\prime |_{h=v_h} = 0$ 
   leads to the relation 
   $|\lambda_{\rm mix}|= 2 \lambda_H (v_h/v_\phi)^2$, 
   and the Higgs boson mass $m_h$ is given by 
\begin{equation}
  m_h^2 = \left. \frac{d^2 V}{dh^2} \right|_{h=v_h} = |\lambda_{\rm mix}|v_\phi^2 = 2 \lambda_H v_h^2. 
\label{Eq:mass_h}
\end{equation}
In the following RG analysis, 
  this is used as the boundary condition for $\lambda_{\rm mix}$ at the renormalization scale $\mu=v_\phi$. 
Note that since $\lambda_H \sim 0.1$ and $v_\phi \gtrsim 10$ TeV by the large electron-positron collider (LEP) constraint \cite{LEP:2003aa, Carena:2004xs, Schael:2013ita},  
  $|\lambda_{\rm mix}| \lesssim 10^{-5}$, which is very small.

In our discussion about the U(1)$^\prime$ symmetry breaking, we neglected $\lambda_{\rm mix}$ 
  by assuming it to be negligibly small. 
Here we justify this treatment. 
In the presence of $\lambda_{\rm mix}$ and the Higgs VEV, Eq.~(\ref{eq:stationary}) is modified as 
\begin{eqnarray}
 \lambda_\Phi = \frac{11}{6} \beta_\Phi + \frac{|\lambda_{\rm mix}|}{2} \left( \frac{v_h}{v_\phi} \right)^2 
 \simeq  
  \frac{1}{2 v_\phi^4}  \left(  \frac{11}{8 \pi^2} m_{Z^\prime}^4 + m_h^2 v_h^2    \right).   
\label{eq:consistency} 
\end{eqnarray}
Considering the current LHC Run-2 bound from search for $Z^\prime$ boson resonances
  \cite{ATLAS:2016cyf, CMS:2016abv},  
  $m_{Z^\prime} \gtrsim 4$ TeV,  we find that the first term in the parenthesis 
  in the last equality is 5 orders of magnitude greater than the second term, 
  and therefore we can analyze the two Higgs sectors separately.

\subsection{Solving the electroweak vacuum instability}

\begin{figure}[t]
\begin{minipage}{0.5\linewidth}
\begin{center}
\includegraphics[width=0.95\linewidth]{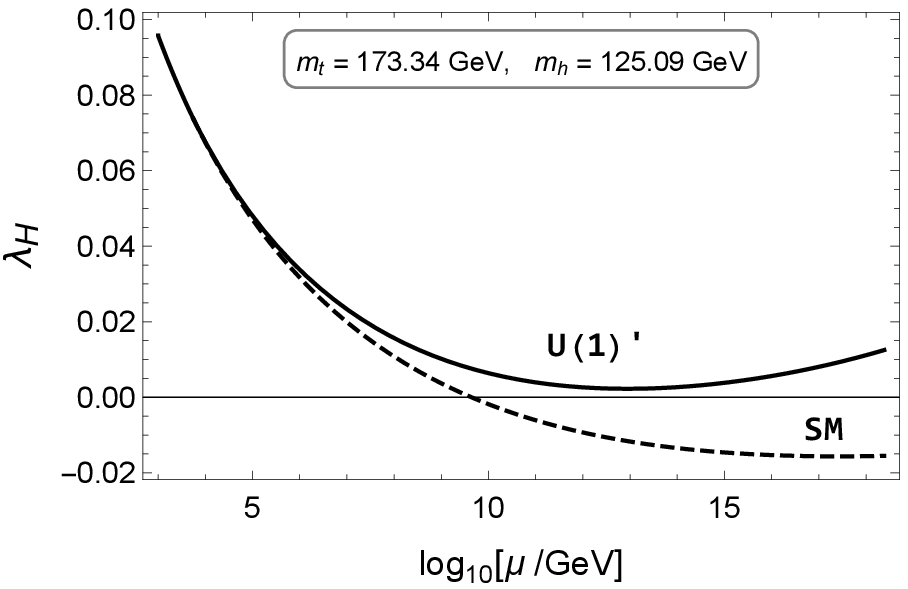}
\subcaption{}\label{Fig:Higgs_stability_lambdaH}
\end{center}
\end{minipage}
\begin{minipage}{0.5\linewidth}
\begin{center}
\includegraphics[width=0.95\linewidth]{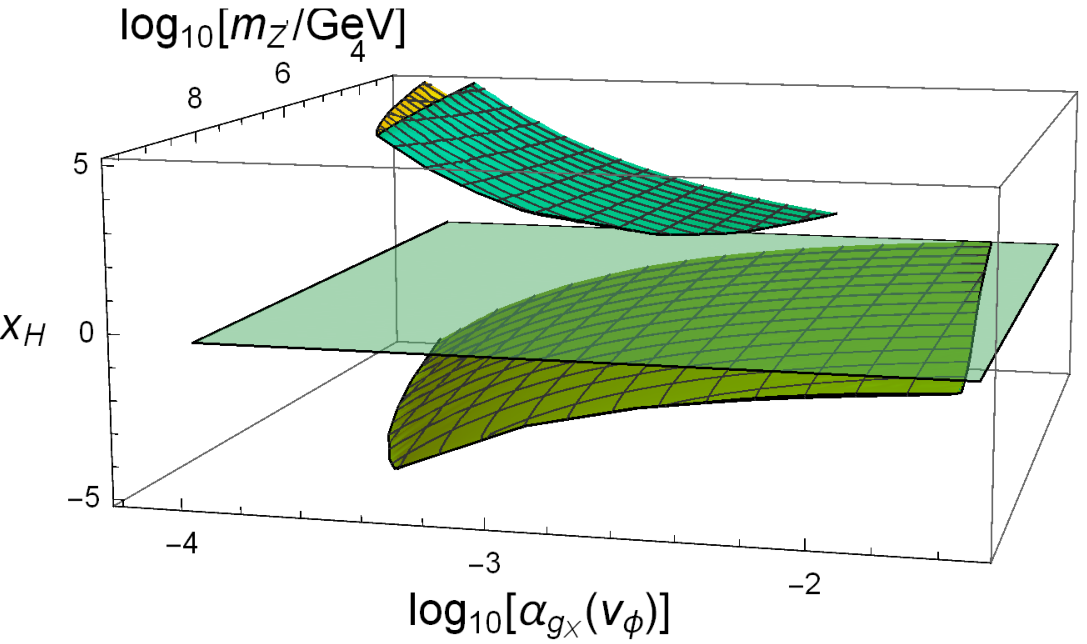}
\subcaption{}\label{Fig:Higgs_stability_3D}
\end{center}
\end{minipage}
\caption
{
\subref{Fig:Higgs_stability_lambdaH} The evolutions of the Higgs quartic coupling $\lambda_H$ (solid line) 
    for the inputs $m_t=173.34$ GeV and $m_h=125.09$ GeV, 
    along with the SM case (dashed line). 
Here, we have taken $x_H =-0.575$, $m_{Z^\prime}=4$ TeV and $\alpha_{g_X}=0.01$,
    which corresponds to $v_\phi = 5.64$ TeV and $g_{X}(v_\phi) = 0.354$. 
\subref{Fig:Higgs_stability_3D} The result of the three-dimensional parameter scans
   for $v_\phi$, $g_{X}$ and $x_H$, 
   shown in ($m_{Z^\prime}/{\rm GeV}, \alpha_{g_X}, x_H$) parameter space
   with $m_{Z^\prime} \simeq x_\Phi g_{X} v_\phi$.
As a reference, a horizontal plane for $x_H=-16/41$ is shown, which corresponds to the orthogonal case.
}
\label{Fig:Higgs_stability}
\end{figure}

In the SM with the observed Higgs boson mass of $m_h=125.09$ GeV~\cite{Aad:2015zhl},  
   the RG evolution of the SM Higgs quartic coupling shows that the running coupling becomes 
   negative at the intermediate scale $\mu \simeq 10^{10}$ GeV \cite{Buttazzo:2013uya} 
   for $m_t=173.34$ GeV~\cite{ATLAS:2014wva}, and hence the electroweak vacuum is unstable.   
In our U(1)$^\prime$ extended SM, however, there is a parameter region
   to solve this electroweak vacuum instability problem~\cite{Oda:2015gna, Das:2016zue}.\footnote{
In the absence of the classical conformal invariance,
   the electroweak vacuum instability problem has been investigated in Refs.~\cite{Coriano:2014mpa,DiChiara:2014wha,
   Coriano:2015sea,Accomando:2016sge}.} 
There are only three free parameters in our model, $x_H$, $v_\phi$, and $g_{X}$,  
   which are also interpreted as $x_H$, $m_{Z^\prime}$, and $\alpha_{g_X}=g_X^2/(4 \pi)$.
Inputs of the couplings at $v_\phi$ are determined by these three parameters.
In Fig.~\ref{Fig:Higgs_stability}\subref{Fig:Higgs_stability_lambdaH},
    we show the RG evolution of the SM Higgs quartic coupling 
    in our model (solid line), along with the SM result (dashed line).  
Here, we have taken $x_H=-0.575$, $m_{Z^\prime}=4$ TeV and $\alpha_{g_X}=0.01$,
    which corresponds to $v_\phi = 5.64$ TeV and $g_{X}(v_\phi) = 0.354$, as an example.  
The Higgs quartic coupling remains positive all the way up to the Planck mass scale, 
   so the electroweak vacuum instability problem is solved.

In order to identify a parameter region to resolve the electroweak vacuum instability, 
   we perform parameter scans for the free parameters $x_H$, $v_\phi$ and $g_{X}$. 
In this analysis, we impose several conditions on the running couplings at $v_\phi \leq \mu \leq M_{\rm P}$ 
   ($M_{\rm P} =2.44 \times 10^{18}$ GeV is the reduced Planck mass): 
   stability conditions of the Higgs potential  ($\lambda_H,  \lambda_\Phi > 0$), 
   and the perturbative conditions that all the running couplings remain in the perturbative regime, namely, 
   $g_i^2$ ($i=1,2,3$), $g_{X}^2$, $g_{X1}^2$, $g_{1X}^2<4\pi$
   and $\lambda_H$, $\lambda_\Phi$, $\lambda_{\rm mix}<4 \pi$. 
For theoretical consistency, we also impose a condition that the 2-loop beta functions
   are smaller than the 1-loop beta functions (see Ref.~\cite{Das:2016zue} for detail). 
In Fig.~\ref{Fig:Higgs_stability}\subref{Fig:Higgs_stability_3D}, we show the result of our parameter scans 
   in the three-dimensional parameter space of ($m_{Z^\prime}, \alpha_{g_X}, x_H$).
As a reference, we show a horizontal plane corresponding to the orthogonal case $x_H=-16/41$.  
There is no overlapping of the plane with the resultant parameter regions 
    to resolve the electroweak vacuum instability.

\subsection{Naturalness bounds from SM Higgs mass corrections}
\label{Sec_naturalness}
Once the classically conformal symmetry is radiatively broken by the CW mechanism,
    the masses for the $Z^\prime$ boson and the Majorana RHNs are generated, 
    and they contribute to self-energy corrections of the SM Higgs doublet. 
If the U(1)$^\prime$ gauge symmetry breaking scale is very large, 
  the self-energy corrections may exceed the electroweak scale
  and require us to fine-tune the model parameters in reproducing the correct electroweak scale.
See \cite{Casas:2004gh} for related discussions. 
As heavy states, we have the RHNs and $Z^\prime$ boson, 
  whose masses are generated by the U(1)$^\prime$ gauge symmetry breaking.

Since the original theory is classically conformal and defined as a massless theory,  
   the self-energy corrections to the SM Higgs doublet originate 
   from corrections to the mixing quartic coupling $\lambda_{\rm mix}$.     
Thus, what we calculate to derive the naturalness bounds are quantum corrections to 
   the term $\lambda_{\rm mix} h^2 \phi^2$ in the effective Higgs potential 
\begin{eqnarray} 
  V_{\rm eff} \supset
    \frac{\lambda_{\rm mix}}{4} h^2 \phi^2 
	+ \frac{\beta_{\lambda_{\rm mix}}}{8} h^2 \phi^2 \left( \ln\left[\phi^2\right] + C \right),
\end{eqnarray}
   where the logarithmic divergence and the terms independent of $\phi$ are all encoded in $C$.
Here, the major contributions to quantum corrections are from the $Z^\prime$ boson loops:
\begin{eqnarray}
\beta_{\lambda_{\rm mix}} &\supset& \frac{12 x_H^2 x_\Phi^2 g_X^4}{16 \pi^2} 
	- \frac{4 \left( 19 x_H^2 + 10 x_H x_\Phi + x_\Phi^2 \right) x_\Phi^2 y_t^2 g_X^4}{\left(16 \pi^2\right)^2},
\end{eqnarray}
  where the first term is from the one-loop diagram,
  and the second one is from the two-loop diagram \cite{Iso:2009ss,Iso:2009nw}
  involving the $Z^\prime$ boson and the top quark.
By adding a counter-term, we renormalize the coupling $\lambda_{\rm mix}$ with the renormalization condition, 
\begin{eqnarray}
   \frac{\partial^4  V_{\rm eff}}{\partial h^2 \partial \phi^2} \Big|_{h=0, \phi=v_\phi} = \lambda_{\rm mix}, 
\end{eqnarray}
   where $\lambda_{\rm mix}$ is the renormalized coupling.
As a result, we obtain 
\begin{eqnarray} 
  V_{\rm eff} \supset
    \frac{\lambda_{\rm mix}}{4} h^2 \phi^2 
	+ \frac{\beta_{\lambda_{\rm mix}}}{8} h^2 \phi^2 \left( \ln\left[\frac{\phi^2}{v_\phi}\right] - 3 \right).
\end{eqnarray}
Substituting $\phi=v_\phi$, we obtain the SM Higgs self-energy correction as 
\begin{eqnarray}    
   \Delta m_h^2 &=& - \frac{3}{4} \beta_{\lambda_{\rm mix}} v_\phi^2 \nonumber \\
	&\sim& - \frac{9}{4 \pi} x_H^2 \alpha_{g_X} m_{Z^\prime}^2
		+ \frac{3 m_t^2} {32 \pi^3 v_h^2} \left(19x_H^2 + 20x_H + 4\right) \alpha_{g_X} m_{Z^\prime}^2.
 \label{Eq:Delta_mh2}
\end{eqnarray}
For the stability of the electroweak vacuum, we impose $\Delta m_h^2 \lesssim m_h^2$ as the naturalness. 
The most important contribution to $\Delta m_h^2$ is the first term of Eq.~(\ref{Eq:Delta_mh2})
  generated through the one-loop diagram with the $Z^\prime$ gauge boson,
  and the second term becomes important in the case of the U(1)$_{B-L}$ model,
  where $x_H=0$.

If $\Delta m_h^2$ is much larger than the electroweak scale,  
  we need a fine-tuning of the tree-level Higgs mass ($|\lambda_{\rm mix}| v_\phi^2/2$) 
  to reproduce the correct SM Higgs VEV, $v_h=246$ GeV. 
We simply evaluate a fine-tuning level as 
\begin{eqnarray}
  \delta  = \frac{m_h^2}{2 |\Delta m_h^2|}. 
\end{eqnarray}
Here, $\delta =0.1$, for example, indicates that we need to fine-tune the tree-level Higgs mass squared 
   at the 10\% accuracy level.

\section{Relic density of the RHN DM}
\label{Sec_relic_density}

In this section, we calculate the thermal relic density of the RHN DM
  and identify the model parameter region to be consistent
  with the Planck 2015 measurement~\cite{Aghanim:2015xee}
  (68\% confidence level):
\begin{eqnarray}
\Omega_{\rm DM} h^2 &=& 0.1198 \pm 0.0015.
\end{eqnarray}
In our model, the RHN DM particles mainly annihilate into the SM particles 
  through the $s$-channel process mediated by the U(1)$^\prime$ gauge boson $Z^\prime$.

\begin{figure}[t]
\begin{center}
\includegraphics[width=0.5\linewidth]{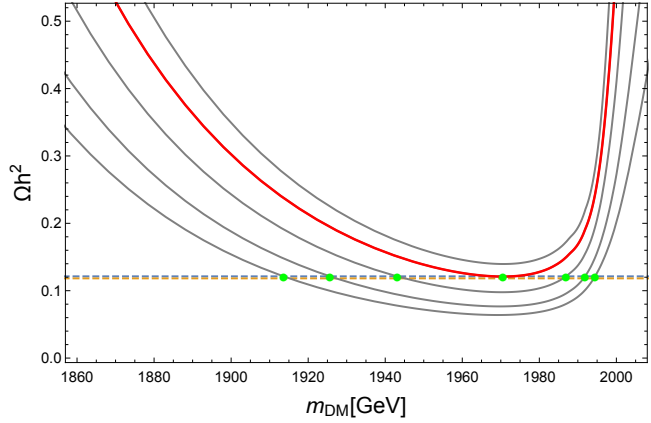}
\caption
{
The relic density of the RHN DM 
  as a function of its mass ($m_{\rm DM}$). 
We have fixed $x_H=-0.575$ and $m_{Z^\prime}=4$ TeV,
  and have shown the relic densities for various values of the gauge coupling,  
  $\alpha_{g_X}=0.002$, $0.00235$, $0.003$, $0.004$ and $0.005$ 
  (solid lines from top to bottom). 
The two horizontal lines denote the range of the observed DM relic density, 
  $0.1183 \leq \Omega_{\rm DM} h^2 \leq 0.1213$ in the Planck 2015 results~\cite{Aghanim:2015xee}.
}
\label{Fig:Omega_h2}
\end{center}
\end{figure}

The Boltzmann equation of the RHN DM is given by 
\begin{eqnarray}
\frac{dY}{dx} &=& -\frac{xs \langle \sigma v \rangle}{H(m_{\rm DM})} (Y^2 - Y_{\rm EQ}^2),
\label{Eq:Boltzmann}
\end{eqnarray}
   where temperature of the Universe is normalized by the mass of the RHN DM
   $x=m_{\rm DM}/T$, 
   $H(m_{\rm DM})$  is the Hubble parameter at $T=m_{\rm DM}$,
   $s$ is the entropy density,
   $Y=n/s$ is the yield of the RHN DM
   which is defined by the ratio of the number density $n$ to $s$,
   $Y_{\rm EQ}$ is the yield in the thermal equilibrium, 
   and $\langle \sigma v \rangle$ is the thermal averaged product of  the RHN DM
   annihilation cross section $\sigma$ and relative velocity $v$.
Explicit formulas of these are summarized as follows:  
\begin{eqnarray}
s &=& \frac{2 \pi^2}{45} g_* \frac{m_{\rm DM}^3}{x^3}, \nonumber \\
H(m_{\rm DM}) &=& \sqrt{\frac{\pi^2}{90} g_*} \frac{m_{\rm DM}^2}{M_{\rm P}},  \nonumber \\
sY_{\rm EQ} &=& \frac{g_{\rm DM}}{2 \pi^2} \frac{m_{\rm DM}^3}{x} K_2(x),  
\end{eqnarray}
   where
   $g_{\rm DM}=2$ is the number of degrees of the freedom for the RHN DM,
   $g_*$ is the effective total number of degrees of freedom for particles in thermal equilibrium
   (in this paper, we set $g_*=106.75$ for the SM particles),
   and  $K_2$ is the modified Bessel function of the second kind.
The thermally-averaged annihilation cross section times velocity is given by
\begin{eqnarray}
\langle \sigma v \rangle 
	&=& (sY_{\rm EQ})^{-2} g_{\rm DM}^2 \frac{m_{\rm DM}}{64 \pi^4 x}
		 \int_{4m_{\rm DM}^2}^{\infty} ds \hat{\sigma}(s) \sqrt{s}
		 K_1\left( \frac{x \sqrt{s}}{m_{\rm DM}} \right),
\end{eqnarray}
   where the reduced cross section is defined as $\hat{\sigma}(s) = 2 (s-4m_{\rm DM}^2)  \sigma(s)$
   with the total cross section $\sigma(s)$,
   $K_1$ is the modified Bessel function of the first kind.
The total cross section of the RHN DM annihilation process
    $\nu_R^3 \nu_R^3 \to Z^\prime \to f \bar{f}$
    ($f$ denotes the SM fermion)\footnote{
Although there are also other annihilation processes, 
   such as $\nu_R^3 \nu_R^3 \to \phi \phi$,
   $\nu_R^3 \nu_R^3 \to \phi Z^\prime$
   and $\nu_R^3 \nu_R^3 \to Z^\prime Z^\prime$
   (see, for example, Ref.~\cite{Bell:2016uhg}),
   all these cross sections are estimated to be much less than 1 pb,
   which is a typical cross section to reproduce
   $\Omega_{\rm DM} h^2 \simeq 0.1$, 
   for $\alpha_{g_X} \sim 0.01$
   (see Figs.~\ref{Fig:mZp_alpha} and \ref{Fig:xH_alpha}),
   $y_{\rm DM} \sim g_X$,
   and $m_{\rm DM} \sim 1$ TeV. 
}
    is calculated as
\begin{eqnarray}
\sigma(s) &=& \frac{\pi}{3} \alpha_{g_X}^2
	\frac{\sqrt{s (s-4m_{\rm DM}^2)}}{(s-m_{Z^\prime}^2)^2+m_{Z^\prime}^2 \Gamma_{Z^\prime}^2} \nonumber \\
	&\times& \left[ \frac{103x_H^2 + 86x_H + 37}{3} 
		+ \frac{17x_H^2 + 10x_H + 2 + (7x_H^2 + 20x_H + 4)\frac{m_t^2}{s}}{3} 
	 		\sqrt{1-\frac{4m_t^2}{s}} \right.  \nonumber \\
	&& \hspace{20mm}
		\left. + 18 x_H^2 \frac{(s-m_{Z^\prime}^2)^2}{s (s-4m_{\rm DM}^2)}
		\frac{m_{\rm DM}^2 m_t^2}{m_{Z^\prime}^4}
		\sqrt{1-\frac{4m_t^2}{s}} \right],
\end{eqnarray}
where the total decay width of $Z^\prime$ boson is given by
\begin{eqnarray}
\Gamma_{Z^\prime} &=& 
	\frac{\alpha_{g_X} m_{Z^\prime}}{6} \!\!
	\left[ \frac{103x_H^2 + 86x_H + 37}{3} 
		+ \frac{17x_H^2 + 10x_H + 2 + (7x_H^2 + 20x_H + 4)\frac{m_t^2}{m_{Z^\prime}^2}}{3} 
	 		\sqrt{1-\frac{4m_t^2}{m_{Z^\prime}^2}} \right.  \nonumber \\
	&& \hspace{20mm}
		\left. + 2 \left( 1 - \frac{4m_N^2}{m_{Z^\prime}^2} \right)^{\frac{3}{2}} 
			\theta \left( \frac{m_{Z^\prime}^2}{m_N^2} -4\right) 
				+  \left( 1 - \frac{4m_{\rm DM}^2}{m_{Z^\prime}^2} \right)^{\frac{3}{2}} 
			\theta \left( \frac{m_{Z^\prime}^2}{m_{\rm DM}^2} -4\right) \right]. 
\label{DecayWidthZp}
\end{eqnarray}
Here, we have neglected all SM fermion masses except for the top quark mass $m_t$.

By solving the Boltzmann Eq.~(\ref{Eq:Boltzmann}) numerically,
  we find the asymptotic value of the yield $Y(\infty)$,
  and the present DM relic density is given by
\begin{eqnarray}
\Omega_{\rm DM} h^2 &=& \frac{m_{\rm DM} s_0 Y(\infty)}{\rho_c / h^2},
\end{eqnarray}
  where $s_0=2890$ cm$^{-3}$ is the entropy density of the present Universe,
  and $\rho_c/h^2=1.05 \times 10^{-5}$ GeV/cm$^3$ is the critical density.
Our analysis involves four parameters, 
   namely $\alpha_{g_X}$, $m_{Z^\prime}$, $m_{\rm DM}$ and $x_H$.       
For $m_{Z^\prime}=4$ TeV and $x_H=-0.575$,
   we show in Fig.~\ref{Fig:Omega_h2} the resultant RHN DM relic density   
   as a function of the RHN DM mass $m_{\rm DM}$, 
   along with the range of the observed DM relic density, 
   $0.1183 \leq \Omega_{\rm DM} h^2 \leq 0.1213$~\cite{Aghanim:2015xee} 
   (two horizontal dashed lines). 
The solid lines from top to bottom show the resultant RHN DM relic densities
   for various values of the gauge coupling, $\alpha_{g_X}=0.002$, $0.00235$, $0.003$, $0.004$ and $0.005$.  
The plots indicate the lower bound on $\alpha_{g_X} \geq 0.00235$  
   for $m_{Z^\prime}=4$ TeV and $x_H=-0.575$ 
   in order to reproduce the observed relic density. 
In addition, we can see that the enhancement of the RHN DM annihilation cross section 
   via the $Z^\prime$ boson resonance is necessary to satisfy the cosmological constraint 
   and hence, $m_{\rm DM} \simeq m_{Z^\prime}/2$.

\section{Collider constraints on the U(1)$^\prime$ $Z^\prime$ boson}
\label{Sec_collider}
\begin{figure}[t]
\begin{minipage}{0.5\linewidth}
\begin{center}
\includegraphics[width=0.95\linewidth]{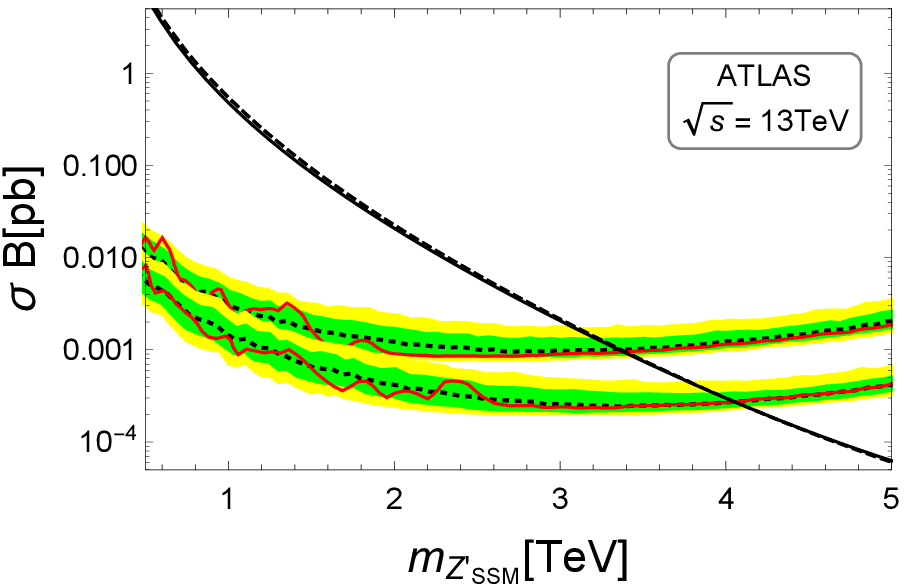}
\subcaption{}\label{Fig:LHC2016_ATLAS}
\end{center}
\end{minipage}
\begin{minipage}{0.5\linewidth}
\begin{center}
\includegraphics[width=0.95\linewidth]{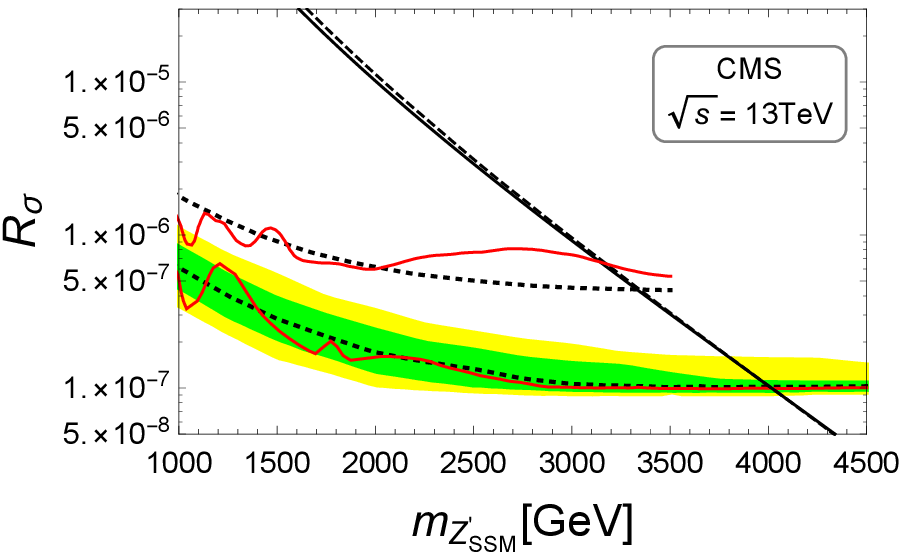}
\subcaption{}\label{Fig:LHC2016_CMS}
\end{center}
\end{minipage}
\caption
{
\subref{Fig:LHC2016_ATLAS} The cross section as a function of the $Z^\prime_{SSM}$ mass (solid line) 
     with $k=1.16$, along with the LHC Run-2 ATLAS result 
     from the combined dielectron and dimuon channels in Ref.~\cite{ATLAS:2016cyf}. 
     (Here we have also shown the ALTAS 2015 result \cite{TheATLAScollaboration:2015jgi}
     for comparison.)
\subref{Fig:LHC2016_CMS} The cross section ratio as a function of the $Z^\prime_{SSM}$ mass (solid line) 
     with $k=1.42$, along with the LHC Run-2 CMS result
     from the combined dielectron and dimuon channels in Ref.~\cite{CMS:2016abv}. 
     (Here we have also shown the CMS 2015 result \cite{CMS:2015nhc} for comparison.)
}
\label{Fig:LHC2016}
\end{figure}
\begin{figure}[t]
\begin{center}
\includegraphics[width=0.5\linewidth]{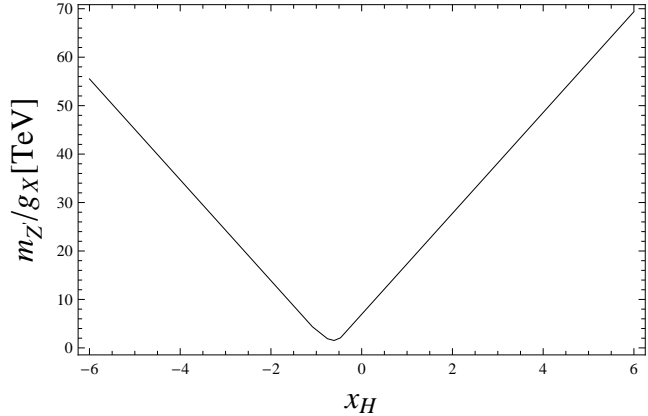}
\caption
{
The lower bound on $m_{Z^\prime}/g_X$ as a function of $x_H$, 
   obtained by the limits from the final LEP 2 data~\cite{Schael:2013ita} 
   at 95\% confidence level. 
}
\label{Fig:LEP}
\end{center}
\end{figure}
The ATLAS and the CMS Collaborations have searched for $Z^\prime$ boson resonance 
  at the LHC Run-1 with $\sqrt{s}=8$ TeV, 
  and continued the search at the LHC Run-2 with $\sqrt{s}=13$ TeV. 
The most stringent bounds on the $Z^\prime$ boson production cross section times branching ratio 
  have been obtained by using the dilepton final state. 
For the so-called sequential SM $Z^\prime$  ($Z^\prime_{SSM}$) model \cite{Barger:1980ti},
  where the $Z^\prime_{SSM}$ boson has exactly 
  the same couplings with the SM fermions as those of the SM $Z$ boson, 
  the latest cross section bounds from the LHC Run-2 results lead to lower bounds
  on the $Z^\prime_{SSM}$ boson mass
  as $m_{Z^\prime_{SSM}} \geq 4.05$ TeV in the ATLAS 2016 results \cite{ATLAS:2016cyf} and 
  $m_{Z^\prime_{SSM}} \geq 4.0$ TeV in the CMS 2016 results \cite{CMS:2016abv}, respectively.  
We interpret these ATLAS and CMS results into the U(1)$^\prime$ $Z^\prime$ boson case  
  and derive constraints on $x_H$, $\alpha_{g_X}$ and $m_{Z^\prime}$.

We calculate the dilepton production cross section
   for the process $pp \to Z^\prime +X \to \ell^{+} \ell^{-} +X$. 
The differential cross section with respect to the invariant mass $M_{\ell \ell}$ of the final state dilepton 
   is described as
\begin{eqnarray}
\frac{d \sigma}{d M_{\ell \ell}}
	= \sum_{a,b}
		\int^{1}_{\frac{M^2_{\ell \ell}}{E^2_{\rm CM}}} d x_1 \frac{2M_{\ell \ell}}{x_1 E^2_{\rm CM}} 
		f_{a}(x_{1}, M^2_{\ell \ell}) f_{b}\left(\frac{M^2_{\ell \ell}}{x_{1} E^2_{\rm CM}}, M^2_{\ell \ell} \right)  
		\hat{\sigma} (\bar{q} q \to Z^\prime \to  \ell^+ \ell^-),
\label{CrossLHC}
\end{eqnarray}
where $f_a$ is the parton distribution function for a parton $a$, 
  and $E_{\rm CM} =13$ TeV is the center-of-mass energy of the LHC Run-2.
In our numerical analysis, we employ CTEQ5M~\cite{Pumplin:2002vw} for the parton distribution functions. 
In the case of the U(1)$^\prime$ model,
  the cross sections for the colliding partons are given by 
\begin{eqnarray}
\hat{\sigma} (\bar{u} u \rightarrow Z^\prime \to \ell^+ \ell^-)  
	&=& \frac{\pi \alpha_{g_X}^2}{81} 
		\frac{M_{\ell \ell}^2}{(M_{\ell \ell}^2-m_{Z^\prime}^2)^2 + m_{Z^\prime}^2 \Gamma_{Z^\prime}^2}
		(85x_H^4 + 152x_H^3 + 104x_H^2 + 32x_H + 4), 
\nonumber\\
\hat{\sigma}  (\bar{d} d \rightarrow Z^\prime \to \ell^+ \ell^-) 
	&=& \frac{\pi \alpha_{g_X}^2}{81} 
		\frac{M_{\ell \ell}^2}{(M_{\ell \ell}^2-m_{Z^\prime}^2)^2 + m_{Z^\prime}^2 \Gamma_{Z^\prime}^2}
		(25x_H^4 + 20x_H^3 + 8x_H^2 + 8x_H + 4), 
\label{CrossLHC2}
\end{eqnarray}
where the total decay width of the $Z^\prime$ boson is given in Eq.~(\ref{DecayWidthZp}).
By integrating the differential cross section over a range of $M_{\ell \ell}$ set by the ATLAS
  and CMS analyses, respectively, we obtain the cross section
  as a function of  $x_H$, $\alpha_{g_X}$ and $m_{Z^\prime}$,
  which are compared with the lower bounds obtained by the ATLAS and CMS Collaborations.

In interpreting the ATLAS and the CMS results for the U(1)$^\prime$ $Z^\prime$ boson,  
   we follow the strategy in \cite{Okada:2016gsh}.
We first analyze the sequential SM $Z^\prime$ model to check the consistency of our analysis 
   with the one by the ATLAS and the CMS Collaborations.  
With the  same couplings as the SM, we calculate the differential cross section of the process
   $pp \to Z^\prime_{SSM}+X \to \ell^+ \ell^- +X$ like Eq.~(\ref{CrossLHC}). 
According to the analysis by the ATLAS Collaboration at the LHC Run-2,  
   we integrate the differential cross section for the range of 
   120 GeV$\leq M_{\ell \ell} \leq 6000$ GeV \cite{ATLAS:2016cyf}
   and obtain the cross section of the dilepton production process 
   as a function of the $Z^\prime_{SSM}$ boson mass.  
Our result is shown as a solid line in Fig.~\ref{Fig:LHC2016}\subref{Fig:LHC2016_ATLAS}, 
  along with the plots presented by the ATLAS Collaboration \cite{ATLAS:2016cyf}
  (Here we have also shown the ALTAS 2015 result \cite{TheATLAScollaboration:2015jgi} for comparison.
   We can see that the ATLAS 2016 result has dramatically improved 
   the bound obtained by the ATLAS 2015 result.).
In Fig.~\ref{Fig:LHC2016}\subref{Fig:LHC2016_ATLAS},
  the experimental upper bounds on the $Z^\prime$ boson production 
   cross section are depicted as the horizontal solid (red) curves. 
The theoretical $Z^\prime$ boson production cross section presented in \cite{ATLAS:2016cyf}
   is shown as the diagonal dashed line, 
   and the lower limit of the $Z^\prime_{SSM}$ boson mass is found to be $4.05$ TeV, 
   which can be read off from the intersection point of the theoretical prediction (diagonal dashed line) 
   and the experimental cross section bound (horizontal lower solid (red) curve).  
In order to take into account the difference of the parton distribution functions
  used in the ATLAS analysis and our analysis, and QCD corrections of the process,
  we have scaled our resultant cross section by a factor $k=1.16$
  in Fig.~\ref{Fig:LHC2016}\subref{Fig:LHC2016_ATLAS},
  with which we can obtain the same lower limit of the $Z^\prime_{SSM}$ boson mass
  as $4.05$ TeV. 
We can see that our result (solid line) in Fig.~\ref{Fig:LHC2016}\subref{Fig:LHC2016_ATLAS} 
  with the factor of $k=1.16$
  is very consistent with the theoretical prediction (diagonal dashed line) presented
  by the ATLAS Collaboration. 
We use this factor in the following analysis for the U(1)$^\prime$ $Z^\prime$ production process,
  when we interpret the ATLAS 2016 result.

We apply the same strategy and compare our results for the $Z^\prime_{SSM}$ model 
   with those in the CMS 2016 results~\cite{CMS:2016abv}.
According to the analysis by the CMS Collaboration,  
   we integrate the differential cross section for the range of 
  $0.95 \; m_{Z^\prime_{SSM}} \leq  M_{\ell \ell} \leq  1.05 \; m_{Z^\prime_{SSM}}$ \cite{CMS:2016abv} 
  and obtain the cross section. 
In the CMS analysis, the limits are set on the ratio of the $Z^\prime_{SSM}$ boson cross section 
   to the $Z/\gamma^*$ cross section:
\begin{eqnarray}
R_{\sigma} &=&
	\frac{\sigma (pp \to Z^\prime+X \to \ell \ell +X)}
		{\sigma (pp \to Z+X \to \ell \ell +X)},
\end{eqnarray}
   where the $Z/\gamma^*$ production cross sections 
   in the mass window of $60$ GeV$\leq  M_{\ell \ell} \leq 120$ GeV 
   are predicted to be $1928$ pb at the LHC Run-2 \cite{CMS:2016abv}.   
Our result for the $Z^\prime_{SSM}$ model is shown as the solid line
   in Fig.~\ref{Fig:LHC2016}\subref{Fig:LHC2016_CMS},  
   along with the plot presented in \cite{CMS:2016abv}  
   (Here we have also shown the CMS 2015 result \cite{CMS:2015nhc} for comparison.
   We can see that the CMS 2016 result has dramatically improved 
   the bound obtained by the CMS 2015 result.).
The analyses in this CMS paper leads to the lower limits of the $Z^\prime_{SSM}$ boson mass
  as $4.0$ TeV, 
  which is read off from the intersection point of the theoretical prediction (diagonal dashed line) 
  and  the experimental cross section bound (horizontal lower solid (red) curve).  
In order to obtain the same lower mass limits, 
   we have scaled our resultant cross section by a factor $k=1.42$
   in Fig.~\ref{Fig:LHC2016}\subref{Fig:LHC2016_CMS}.   
With this $k$ factor, our result (solid line) is very consistent
   with the theoretical prediction (diagonal dashed line) presented 
   in Ref.~\cite{CMS:2016abv}.    
We use this $k$ factor in our analysis to interpret the CMS result 
    for the U(1)$^\prime$ $Z^\prime$ boson case.

The search for effective 4-Fermi interactions mediated by the $Z^\prime$ boson at the LEP
  leads to a lower bound on $m_{Z^\prime}/g_X$~\cite{LEP:2003aa,Carena:2004xs,Schael:2013ita}.
Employing the limits from the final LEP 2 data~\cite{Schael:2013ita} at 95\% confidence level, 
   we follow Ref.~\cite{Carena:2004xs} and derive a lower bound on $m_{Z^\prime}/g_X$ 
  as a function $x_H$. 
Our result is shown in Fig.~\ref{Fig:LEP}.

\section{Combined results}
\label{Sec_allowed_region}
\begin{figure}[t]
\begin{center}
\includegraphics[width=0.5\linewidth]{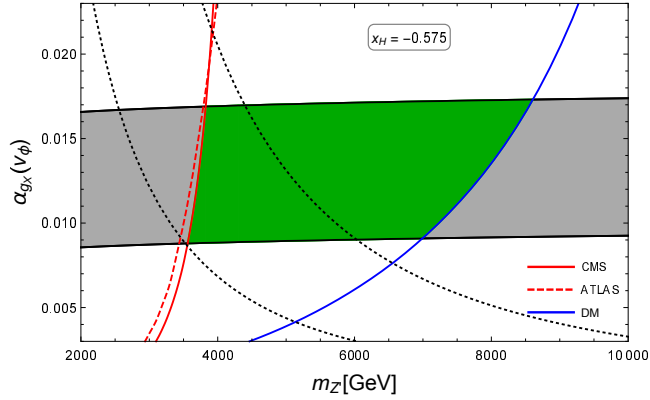}
\caption
{
The allowed regions to solve the electroweak instability problem
    for $m_{Z^\prime}$ and $\alpha_{g_{X}}$ with a fixed $x_H=-0.575$    
    at the TeV scale,
    along with the dark matter lower bound ((blue) right solid line) on $\alpha_{g_{X}}$, 
    the LHC Run-2 (2016) CMS upper bound ((red) solid line) on $\alpha_{g_{X}}$ 
    and the LHC Run-2 ATLAS (2016) upper bound ((red) dashed line) on $\alpha_{g_{X}}$
    from direct search for $Z^\prime$ boson resonance. 
The (green) shaded region in between two solid lines satisfies all constraints.
Here, the naturalness bounds for 10\% (right dotted line) and 30\% (left dotted line) fine-tuning levels
    are also depicted.
}
\label{Fig:mZp_alpha}
\end{center}
\end{figure}
\begin{figure}[htbp]
\begin{minipage}{0.5\linewidth}
\begin{center}
\includegraphics[width=0.95\linewidth]{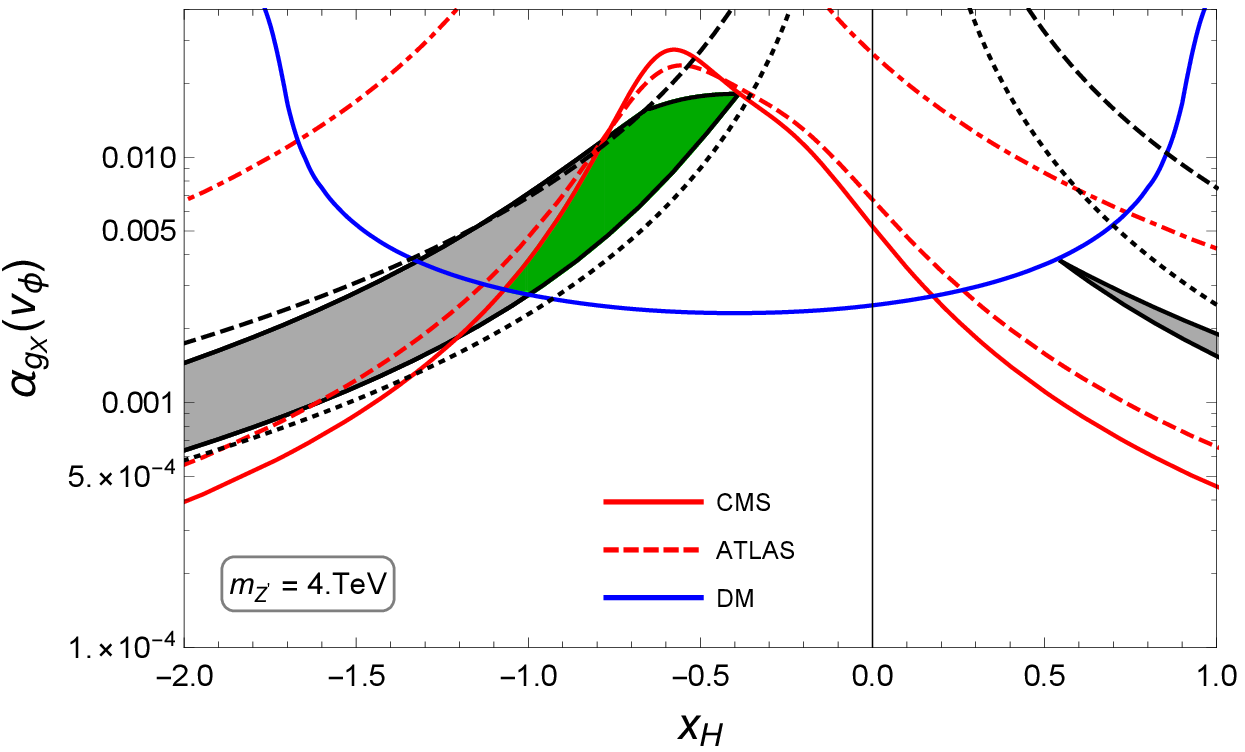}
\subcaption{}\label{Fig:xH_alpha_mzp=4000}
\vspace{5mm}
\end{center}
\end{minipage}
\begin{minipage}{0.5\linewidth}
\begin{center}
\includegraphics[width=0.95\linewidth]{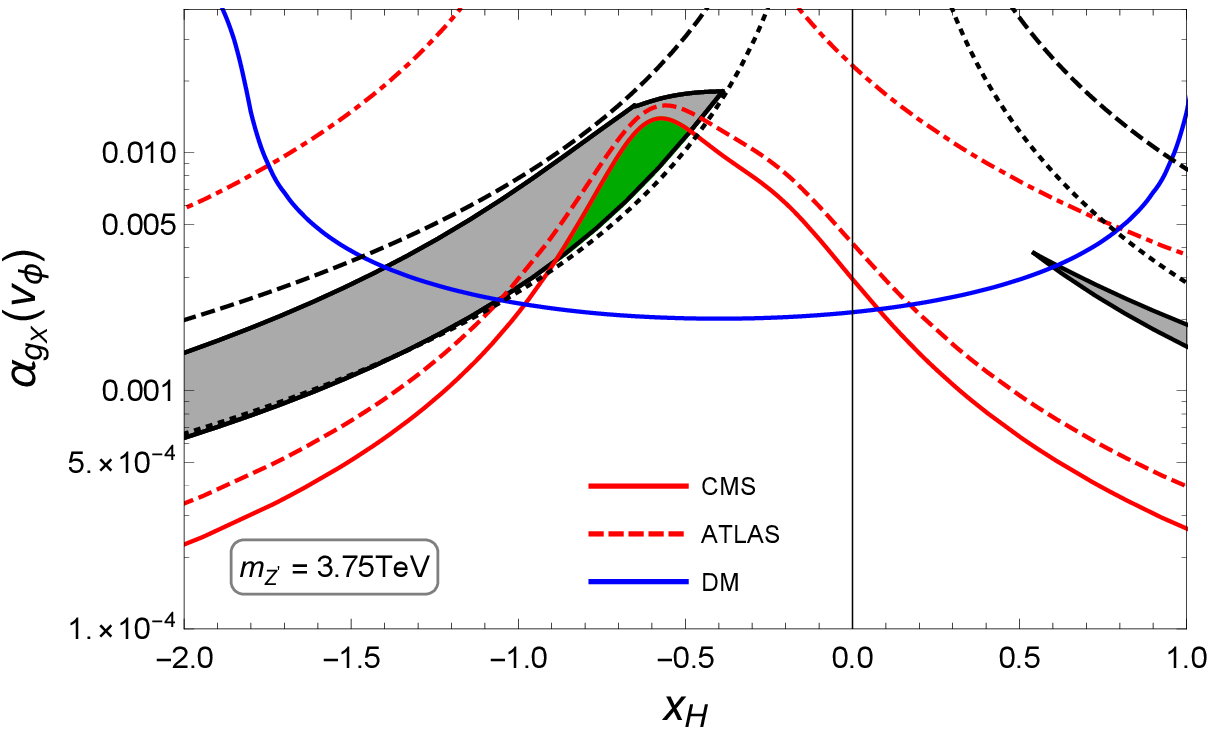}
\subcaption{}\label{Fig:xH_alpha_mzp=3750}
\vspace{5mm}
\end{center}
\end{minipage}
\begin{minipage}{0.5\linewidth}
\begin{center}
\includegraphics[width=0.95\linewidth]{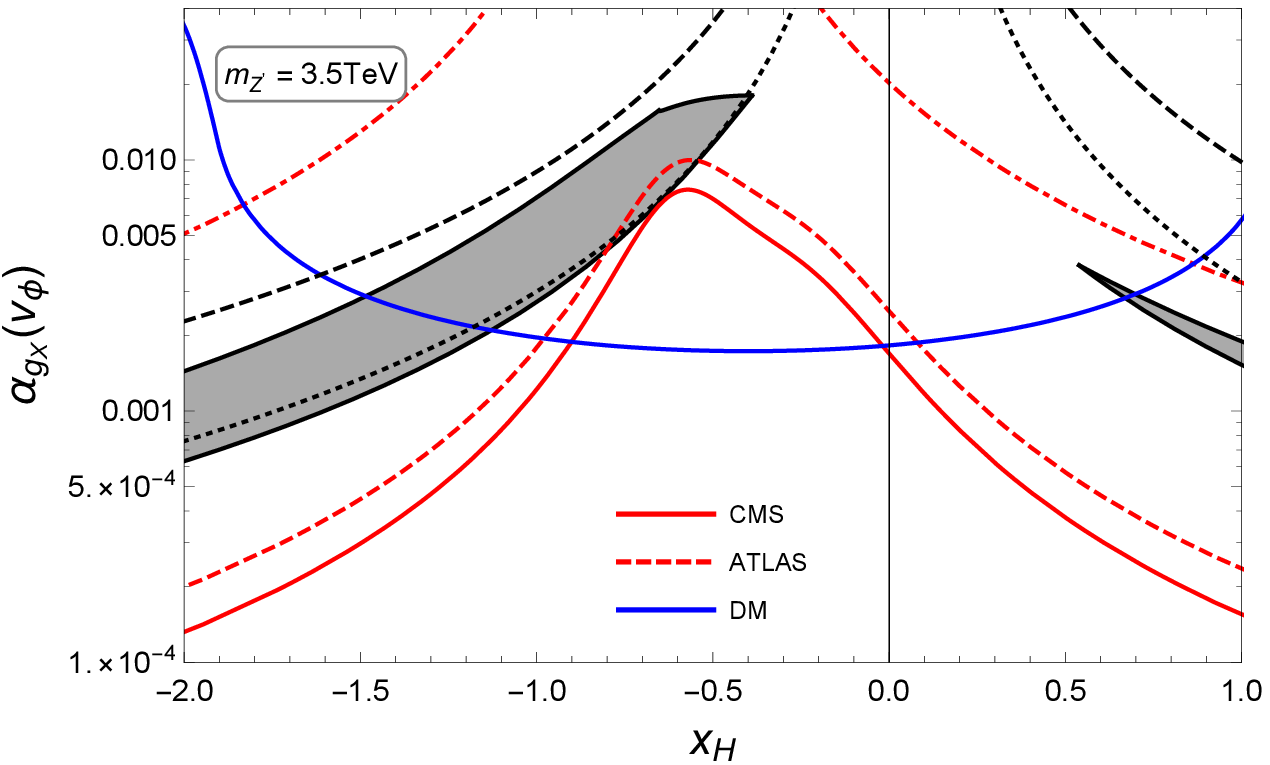}
\subcaption{}\label{Fig:xH_alpha_mzp=3500}
\end{center}
\end{minipage}
\begin{minipage}{0.5\linewidth}
\begin{center}
\includegraphics[width=0.95\linewidth]{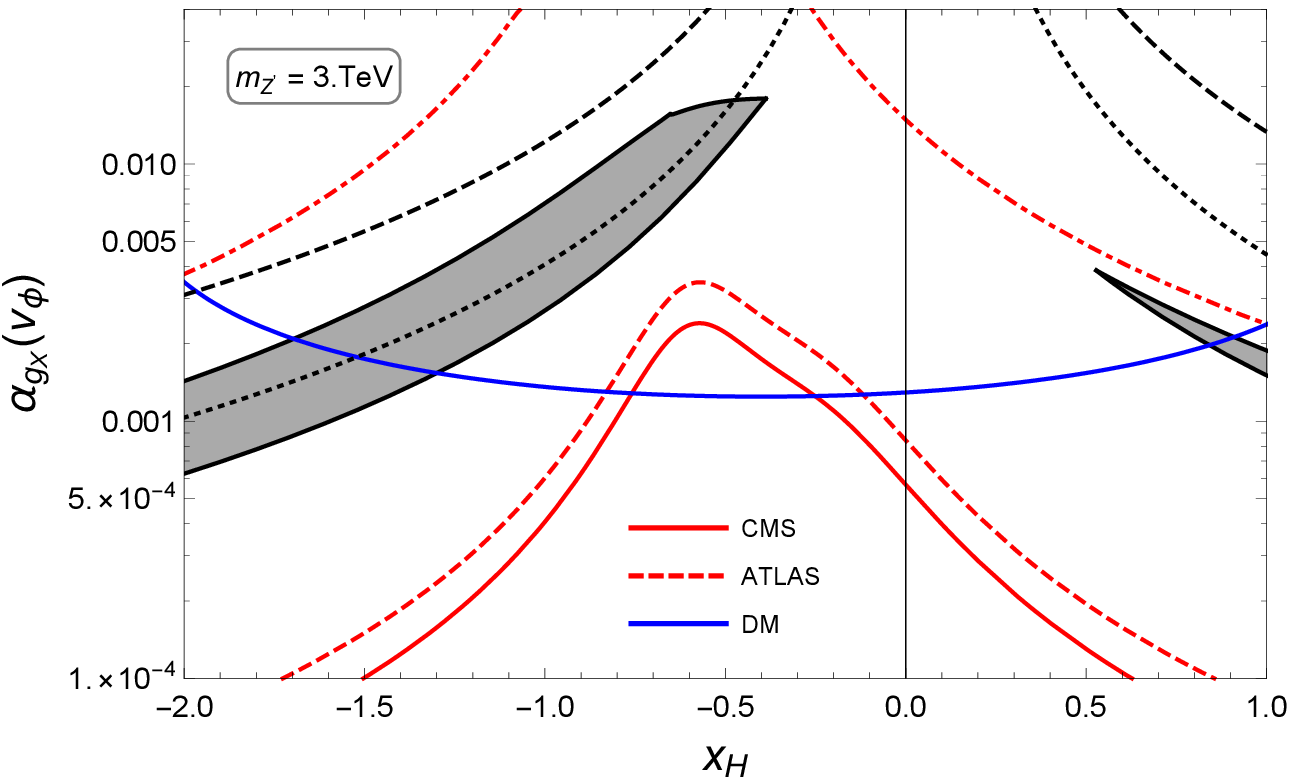}
\subcaption{}\label{Fig:xH_alpha_mzp=3000}
\end{center}
\end{minipage}
\caption{
Allowed parameter regions in the ($x_H$, $\alpha_{g_X}$)-plain 
   for various $m_{Z^\prime}$ values. 
\subref{Fig:xH_alpha_mzp=4000} is for $m_{Z^\prime}=4$ TeV. 
The shaded region indicates the parameter space for solving the electroweak vacuum instability.
The (blue) convex-downward solid line shows the cosmological lower bound on $\alpha_{g_X}$ as a function of $x_H$.
The (red) convex-upward solid (dashed) line shows the upper bound on $\alpha_{g_X}$ 
   obtained from the $Z^\prime$ boson search 
   by the CMS~\cite{CMS:2016abv} (ATLAS~\cite{ATLAS:2016cyf}) Collaboration,
   and the (red) dashed-dotted lines show the LEP bounds. 
The (green) shaded region in between two solid lines satisfies all constraints.
Here, the naturalness bounds for 10\% (dashed line) and 30\% (dotted line) fine-tuning levels
    are also depicted.
\subref{Fig:xH_alpha_mzp=3750}, \subref{Fig:xH_alpha_mzp=3500}
   and \subref{Fig:xH_alpha_mzp=3000} are the same as \subref{Fig:xH_alpha_mzp=4000},
   but $m_{Z^\prime}=3.75$ TeV, 3.5 TeV and 3 TeV, respectively.
}
\label{Fig:xH_alpha}
\end{figure}
Now let us combine all the constraints that we have obtained in the previous sections
   from the RHN DM physics, collider phenomenology, and the electroweak vacuum stability. 
In Fig.~\ref{Fig:mZp_alpha}, we show the allowed region in the ($m_{Z^\prime}$, $\alpha_{g_X}$)-plain 
   for fixed $x_H=-0.575$, as an example. 
The shaded region indicates the parameter space for solving the electroweak vacuum instability. 
The (blue) right solid line shows the lower bound on $\alpha_{g_X}$ as a function of $m_{Z^\prime}$ 
    to reproduce the observed DM relic density of the Planck result~\cite{Aghanim:2015xee}. 
The (red) left solid (dashed) line shows the upper bound on $\alpha_{g_X}$ 
   obtained from the search results for $Z^\prime$ boson resonance 
   by the CMS~\cite{CMS:2016abv} (ATLAS~\cite{ATLAS:2016cyf}) Collaboration.
The (green) shaded region in between two solid lines satisfies all constraints.
These three constraints are complementary to narrow down the allowed region
   to be 4 TeV $\lesssim m_{Z^\prime} \lesssim 8$ TeV and 
   $0.009 \lesssim \alpha_{g_X} \lesssim 0.017$.
We also show the naturalness bounds for 10\% (right dotted line)
   and 30\% (left dotted line) fine-tuning levels.

In Fig.~\ref{Fig:xH_alpha}, we show allowed parameter regions in the ($x_H$, $\alpha_{g_X}$)-plain 
   for various $m_{Z^\prime}$ values. 
Fig.~\ref{Fig:xH_alpha}\subref{Fig:xH_alpha_mzp=4000} is for $m_{Z^\prime}=4$ TeV. 
The shaded region indicates the parameter space for solving the electroweak vacuum instability.
The (blue) convex-downward solid line shows the lower bound on $\alpha_{g_X}$ as a function of $x_H$ 
    to reproduce the observed DM relic density. 
The (red) convex-upward solid (dashed) line shows the upper bound on $\alpha_{g_X}$ 
   obtained from the search results for $Z^\prime$ boson resonance 
   by the CMS~\cite{CMS:2016abv} (ATLAS~\cite{ATLAS:2016cyf}) Collaboration,
   and the (red) dashed-dotted lines also show the LEP bounds. 
The (green) shaded region in between two solid lines satisfies all constraints.
These three constraints are complementary to narrow down the allowed region
   to be $-1.1 \lesssim x_H \lesssim -0.4$ and 
   $0.002 \lesssim \alpha_{g_X} \lesssim 0.02$.
We also show the naturalness bounds for 10\% (dashed line) and 30\% (dotted line) fine-tuning levels.
Figs.~\ref{Fig:xH_alpha}\subref{Fig:xH_alpha_mzp=3750},
   \ref{Fig:xH_alpha}\subref{Fig:xH_alpha_mzp=3500}
   and \ref{Fig:xH_alpha}\subref{Fig:xH_alpha_mzp=3000}
   are the same as Fig.~\ref{Fig:xH_alpha}\subref{Fig:xH_alpha_mzp=4000},
   but $m_{Z^\prime}=3.75$ TeV, 3.5 TeV and 3 TeV, respectively.
From Fig.~\ref{Fig:xH_alpha}\subref{Fig:xH_alpha_mzp=3750},
   the allowed region to satisfy these three constraints indicates
   $-0.9 \lesssim x_H \lesssim -0.5$ and $0.003 \lesssim \alpha_{g_X} \lesssim 0.015$
   for fixed $m_{Z^\prime}=3.75$ TeV.
As $m_{Z^\prime}$ decreases, the LHC upper bound lines are shifted downward,
   while the DM lower bound line remains almost the same (it slightly moves to downward).
Therefore, the allowed region between the LHC upper bounds and the DM lower bound narrows.
On the other hand, the shaded region remains almost the same, 
  so that the (green) shaded region disappears for $m_{Z^\prime} \lesssim 3.5$ TeV.

\section{Direct detection of RHN DM}
\label{Sec_direct_detection}
\begin{figure}[t]
\begin{center}
\includegraphics[width=0.5\linewidth]{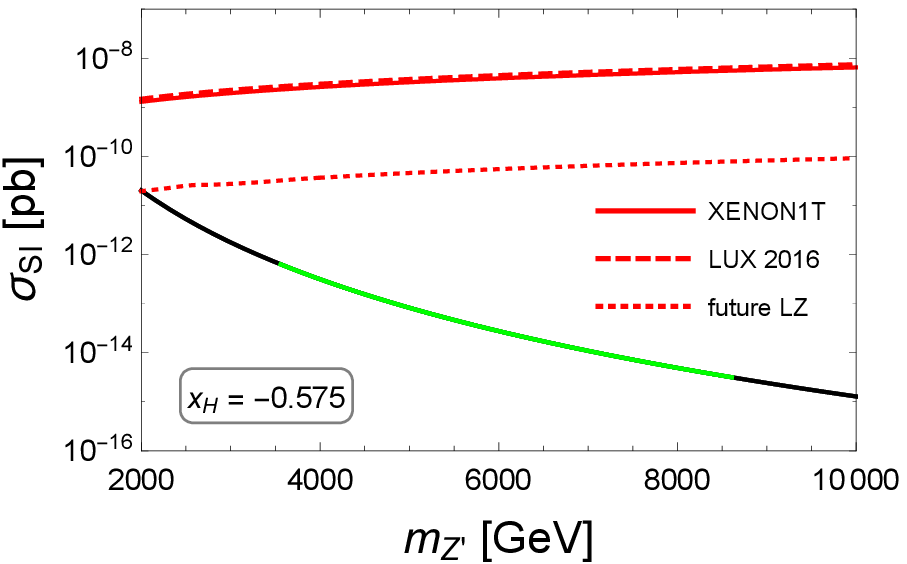}
\caption
{
For a fixed $x_H=-0.575$,
   the resultant spin-independent cross section $\sigma_{\rm SI}$
   as a function of $m_{Z^\prime}$.
Here, for a fixed $m_{Z^\prime}$ value,
   $\alpha_{g_X}$ is taken from the shaded region in Fig.~\ref{Fig:mZp_alpha}
   to solve the electroweak vacuum instability problem.    
The (green) shaded region in between around 3.5 TeV and 9 TeV
   corresponds to the (green) shaded parameter region in Fig.~\ref{Fig:mZp_alpha},
   which satisfies all three constraints,
   the electroweak vacuum stability condition, the LHC Run-2 bound,
   and the cosmological constraint from the observed RHN DM relic density.
The (red) upper solid (dashed) line shows the XENON1T~\cite{Aprile:2017iyp}
   (LUX 2016~\cite{Akerib:2016vxi}) upper bound on $\sigma_{\rm SI}$
   as a function of $m_{Z^\prime} \simeq 2 m_{\rm DM}$,
   and the (red) dotted line shows the prospective reach for the upper bound on $\sigma_{\rm SI}$
   in the next-generation successor of the LUX experiment,
   the LUX-ZEPLIN (LZ) DM experiment~\cite{Szydagis:2016few}.
}
\label{Fig:mZp_sigma}
\end{center}
\end{figure}
\begin{figure}[htbp]
\begin{minipage}{0.5\linewidth}
\begin{center}
\includegraphics[width=0.95\linewidth]{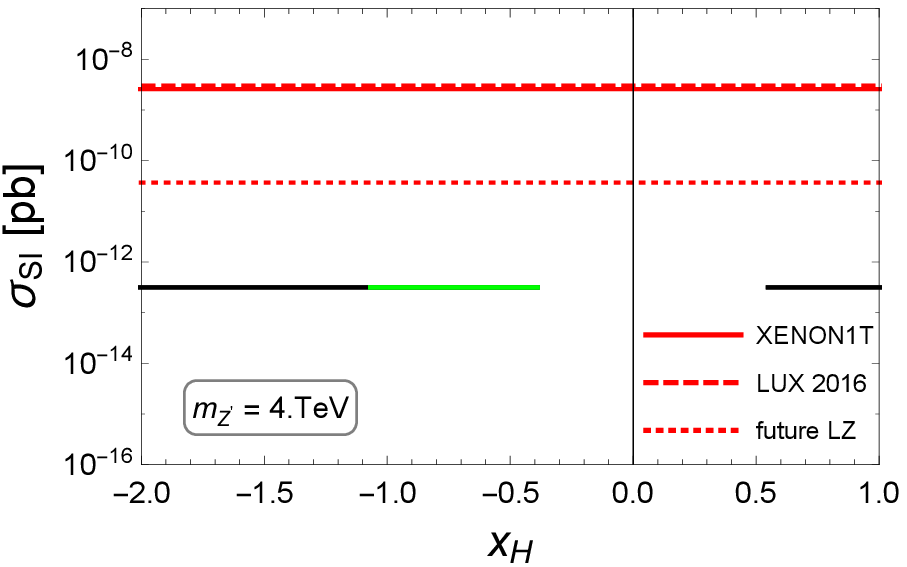}
\subcaption{}\label{Fig:xH_sigma_mzp=4000}
\vspace{5mm}
\end{center}
\end{minipage}
\begin{minipage}{0.5\linewidth}
\begin{center}
\includegraphics[width=0.95\linewidth]{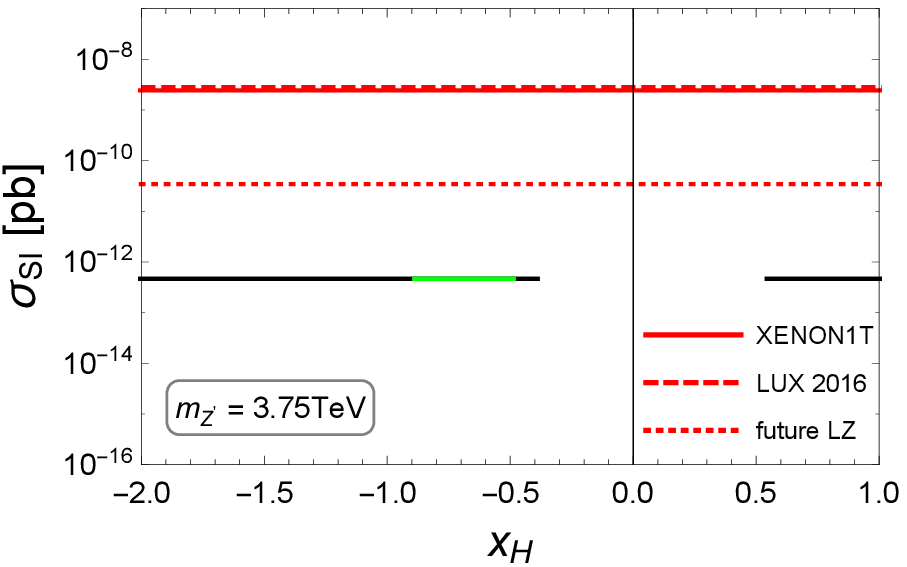}
\subcaption{}\label{Fig:xH_sigma_mzp=3750}
\vspace{5mm}
\end{center}
\end{minipage}
\begin{minipage}{0.5\linewidth}
\begin{center}
\includegraphics[width=0.95\linewidth]{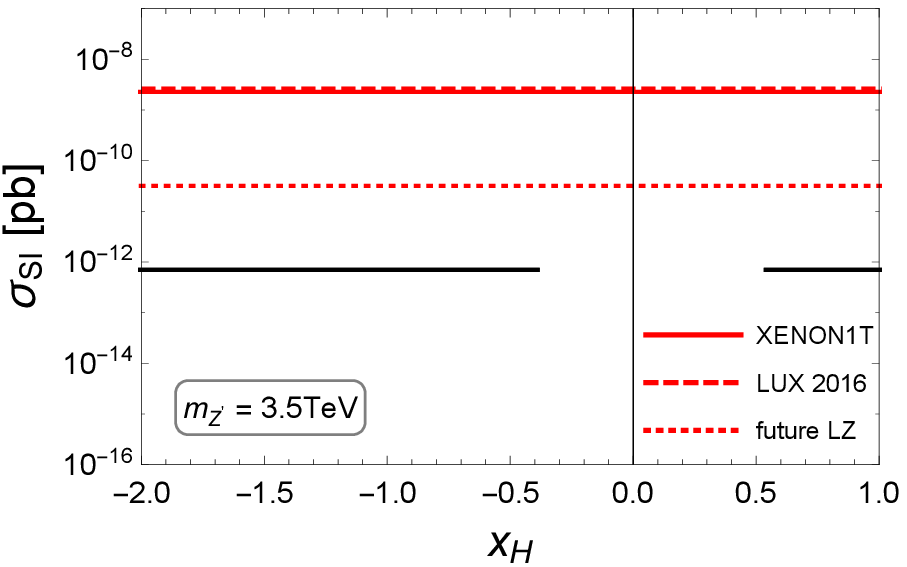}
\subcaption{}\label{Fig:xH_sigma_mzp=3500}
\end{center}
\end{minipage}
\begin{minipage}{0.5\linewidth}
\begin{center}
\includegraphics[width=0.95\linewidth]{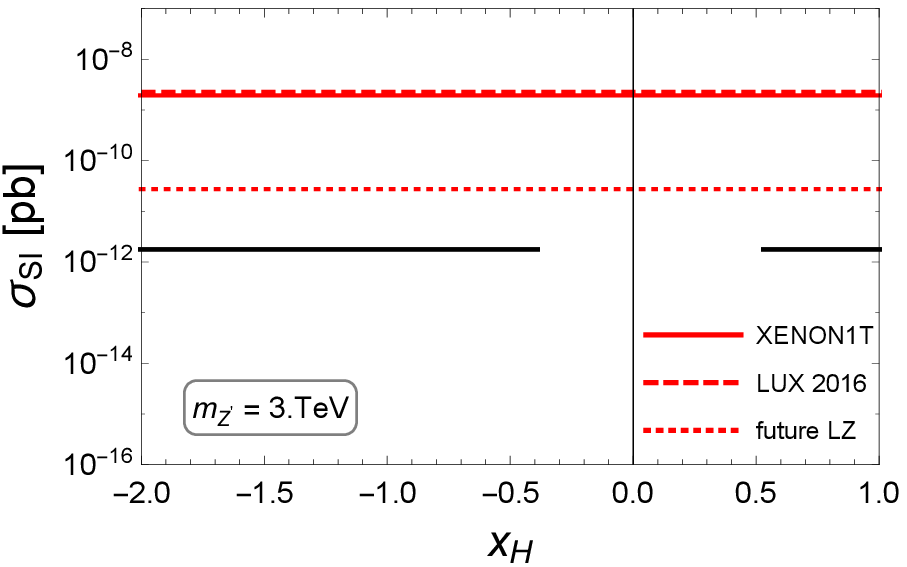}
\subcaption{}\label{Fig:xH_sigma_mzp=3000}
\end{center}
\end{minipage}
\caption{
The resultant $\sigma_{\rm SI}$ in the ($x_H$, $\sigma_{\rm SI}$)-plain 
   for various $m_{Z^\prime}$ values,
   corresponding to the parameter regions shown in Fig.~\ref{Fig:xH_alpha}. 
\subref{Fig:xH_sigma_mzp=4000}  shows our results for $m_{Z^\prime}=4$ TeV. 
The shaded regions indicate the parameter space for solving the electroweak vacuum instability.
The (green) shaded region in the range of $-1.1 \lesssim x_H \lesssim -0.4$
   corresponds to the (green) shaded region
   in Fig.~\ref{Fig:xH_alpha}\subref{Fig:xH_alpha_mzp=4000},
   which satisfies all three constraints,
   the electroweak vacuum stability condition, the LHC Run-2 bound,
   and the cosmological constraint from the observed RHN DM relic density.
The (red) upper solid (dashed) line shows the XENON1T~\cite{Aprile:2017iyp}
   (LUX 2016~\cite{Akerib:2016vxi}) upper bound on $\sigma_{\rm SI}$,
   and the (red) dotted line shows the prospective reach for the upper bound on $\sigma_{\rm SI}$
   in the LZ DM experiment~\cite{Szydagis:2016few}.
Figs.~\subref{Fig:xH_sigma_mzp=3750}, \subref{Fig:xH_sigma_mzp=3500}
   and \subref{Fig:xH_sigma_mzp=3000}
   are the same as \subref{Fig:xH_sigma_mzp=4000},
   but for $m_{Z^\prime}=3.75$ TeV, 3.5 TeV and 3 TeV
   corresponding to Fig.~\ref{Fig:xH_alpha}\subref{Fig:xH_alpha_mzp=3750},
   \ref{Fig:xH_alpha}\subref{Fig:xH_alpha_mzp=3500}
   and \ref{Fig:xH_alpha}\subref{Fig:xH_alpha_mzp=3000}, respectively.
}
\label{Fig:xH_sigma}
\end{figure}
A variety of experiments are underway and also planned for directly detecting 
  a DM particle through its elastic scattering off with nuclei.\footnote{
We can also consider an indirect detection
  of the RHN DM through cosmic rays from
  a pair annihilation of the RHN DMs.
However, using the parameters in the allowed regions
  shown in Sec.~\ref{Sec_allowed_region},
  we have found that the pair annihilation cross section
  is much smaller than the current upper bounds
  obtained from, for example,
  the Fermi-LAT experiments \cite{Charles:2016pgz}.
} 
In this section, we calculate the spin-independent elastic scattering cross section 
  of the RHN DM particle via the Higgs bosons exchange,\footnote{
There is another process for the RHN DM to scatter off with nuclei 
  via $Z^\prime$-boson exchange.
Since the RHN DM is a Majorana particle, 
  only its interaction with nuclei is spin-dependent. 
We have calculated this spin-dependent cross section
  to be  $\sigma_{\rm SD} \sim 10^{-9}$ pb, 
  which is far below the current upper bounds,
  $\sigma_{\rm SD} \lesssim 10^{-4}$ pb 
  obtained from the LUX~\cite{daSilva:2017swg}
  and the IceCube~\cite{Aartsen:2016zhm} experiments.
}
  and compare our results with the current experimental results
  and a prospective reach by future experiments.

From Eq.~(\ref{Eq:mass_Zp_DM}), 
   the U(1)$^\prime$ Higgs VEV $v_\phi$ is expressed
   as a function of $m_{Z^\prime}$, $ \alpha_{g_{X}}$ and $x_H$:
\begin{eqnarray}
v_\phi^2 &=& \frac{m_{Z^\prime}^2}{16 \pi \alpha_{g_{X}}}
		\left[1 - 4 \pi \alpha_{g_{X}} \left( \frac{x_H v_h}{m_{Z^\prime}} \right)^2 \right]
	\; \simeq \; \frac{m_{Z^\prime}^2}{16 \pi \alpha_{g_{X}}}. 
\label{Eq:v_phi2}
\end{eqnarray} 
In Sec.~\ref{Sec_relic_density}, we have also shown that $m_{\rm DM} \simeq m_{Z^\prime}/2$
  to satisfy the experimental relic density of the $Z^\prime$-portal RHN DM,
  which means $y_{\rm DM} \simeq m_{Z^\prime}/2\sqrt{2} v_\phi \simeq \sqrt{2 \pi \alpha_{g_X}}$,
  and then Eq.~(\ref{Eq:mass_phi}) is approximately expressed as
\begin{eqnarray}
m_\phi^2  &\simeq& \frac{1}{8 \pi^2} \frac{23}{8} \frac{m_{Z^\prime}^4}{v_\phi^2}
	\; \simeq \; \frac{23}{4 \pi}  \alpha_{g_X} m_{Z^\prime}^2.
\label{Eq:m_phi2_sim}
\end{eqnarray} 
Using the SM Higgs boson mass in Eq.~(\ref{Eq:mass_h}),
  the scalar mass matrix is found to be 
\begin{eqnarray}
{\cal M} &=& 
	\left(
	\begin{array}{cc}
	 m_h^2 & -m_h^2 \left( \frac{v_h}{v_\phi} \right) \\
	 -m_h^2 \left( \frac{v_h}{v_\phi} \right) & m_\phi^2 
	\end{array}\right).  
\end{eqnarray}
The mass eigenstates $h^\prime$ and $\phi^\prime$ are defined as
\begin{eqnarray}
\left( \begin{array}{c} h^\prime \\ \phi^\prime \end{array} \right)
	&=& \left(
		\begin{array}{cc}
		 \cos \theta & - \sin \theta \\
		 \sin \theta & \cos \theta 
		\end{array}\right)
		\left( \begin{array}{c} h \\ \phi \end{array} \right),
\end{eqnarray}
  with the mixing angle $\theta$ given by 
\begin{eqnarray}
 \tan 2 \theta &=&
	\frac{2m_h^2(v_h/v_\phi)}{m_h^2-m_\phi^2},
\label{Eq:tan_2theta}
\end{eqnarray}
  and their mass eigenvalues are given by
\begin{eqnarray}
m_{h^\prime}^2
	 &=& m_h^2 \cos^2 \theta + m_\phi^2 \sin^2 \theta 
		+ 2m_h^2 \frac{v_h}{v_\phi} \sin \theta \cos \theta
	\; \simeq \; m_h^2, \nonumber \\
m_{\phi^\prime}^2
	 &=& m_h^2 \sin^2 \theta + m_\phi^2 \cos^2 \theta 
		- 2m_h^2 \frac{v_h}{v_\phi} \sin \theta \cos \theta
	\; \simeq \; m_\phi^2.
\label{Eq:m_scalar2}
\end{eqnarray}
Here, we have used the fact that except for the special case, 
  $m_h^2 \simeq m_\phi^2$, 
  the mixing angle is always small because of the suppression 
  by $v_h/v_\phi$ with $v_h=246$ GeV and $v_\phi \gtrsim 10$ TeV. 
Thus, the mass eigenstate $h^\prime$ is the SM-like Higgs boson, 
  while $\phi^\prime$ is the U(1)$^\prime$-like Higgs boson.

The spin-independent elastic scattering cross section with nucleon is given by 
\begin{eqnarray}
\sigma_{\rm SI} 
	&=& \frac{1}{\pi} \left( \sqrt{2} y_{\rm DM} \sin \theta \cos \theta \right)^2
		\left( \frac{\mu_{\rm DM,N}}{v_h} \right)^2 f_N^2 
		\left( \frac{1}{m_{h^\prime}^2} - \frac{1}{m_{\phi^\prime}^2} \right)^2 \nonumber \\
	&\simeq& 4 \theta^2 \alpha_{g_X} 
		\left( \frac{\mu_{\rm DM,N}}{v_h} \right)^2 f_N^2
		\left( \frac{1}{m_h^2} - \frac{1}{m_\phi^2} \right)^2 ,
\label{DD}
\end{eqnarray}
   where $\mu_{\rm DM,N} = m_N m_{{\rm DM}}/(m_N+m_{{\rm DM}})$ is 
   the reduced mass of the RHN DM-nucleon system with the nucleon mass $m_N=0.939$ GeV, and 
\begin{eqnarray}
f_N &=& \left( \sum_{q=u,d,s} f_{T_q} + \frac{2}{9}f_{TG} \right) m_N
\end{eqnarray}
   is the nuclear matrix element accounting for the quark and gluon contents of the nucleon. 
In evaluating $f_{T_q}$, we use the results from the lattice QCD simulation \cite{Ohki:2008ff}: 
   $f_{T_u} +f_{T_d} \simeq 0.056$ and $|f_{T_s}|\leq 0.08$. 
For conservative analysis, we take $f_{T_s}=0$ in the following.     
Using the trace anomaly formula, $\sum_{q=u,d,s} f_{T_q} + f_{TG}=1$
   \cite{Crewther:1972kn,Chanowitz:1972vd,Chanowitz:1972da,Collins:1976yq,Shifman:1978zn}, 
   we obtain $f_N^2 \simeq 0.0706 \; m_N^2$. 
Using Eqs.~(\ref{Eq:v_phi2}), (\ref{Eq:m_phi2_sim}) and (\ref{Eq:tan_2theta}),
   $\sigma_{\rm SI}$ is expressed as a function of only two free parameters:
   $\alpha_{g_X}$ and $m_{Z^\prime}$.

For a fixed $x_H=-0.575$,
   the resultant spin-independent cross section $\sigma_{\rm SI}$
   as a function of $m_{Z^\prime}$
   is depicted in Fig.~\ref{Fig:mZp_sigma}. 
Here, for a fixed $m_{Z^\prime}$ value,
   $\alpha_{g_X}$ is taken from the shaded region in Fig.~\ref{Fig:mZp_alpha}
   to solve the electroweak vacuum instability problem.    
The (green) shaded region in between around 3.5 TeV and 9 TeV
   corresponds to the (green) shaded parameter region in Fig.~\ref{Fig:mZp_alpha},
   which satisfies all three constraints,
   the electroweak vacuum stability condition, the LHC Run-2 bound,
   and the cosmological constraint from the observed RHN DM relic density.
The (red) upper solid (dashed) line shows the XENON1T~\cite{Aprile:2017iyp}
   (LUX 2016~\cite{Akerib:2016vxi}) upper bound on $\sigma_{\rm SI}$
   as a function of $m_{Z^\prime} \simeq 2 m_{\rm DM}$,
   and the (red) dotted line shows the prospective reach for the upper bound on $\sigma_{\rm SI}$
   in the next-generation successor of the LUX experiment,
   the LUX-ZEPLIN (LZ) DM experiment~\cite{Szydagis:2016few}.
Our resultant spin-independent cross section appears below the future reach.

In Fig.~\ref{Fig:xH_sigma}, we show the resultant $\sigma_{\rm SI}$
   in the ($x_H$, $\sigma_{\rm SI}$)-plain for various $m_{Z^\prime}$ values,
   corresponding to the parameter regions shown in Fig.~\ref{Fig:xH_alpha}. 
Fig.~\ref{Fig:xH_sigma}\subref{Fig:xH_sigma_mzp=4000} shows our results for $m_{Z^\prime}=4$ TeV. 
The shaded regions indicate the parameter space for solving the electroweak vacuum instability.
The (green) shaded region in the range of $-1.1 \lesssim x_H \lesssim -0.4$
   corresponds to the (green) shaded region
   in Fig.~\ref{Fig:xH_alpha}\subref{Fig:xH_alpha_mzp=4000},
   which satisfies all three constraints,
   the electroweak vacuum stability condition, the LHC Run-2 bound,
   and the cosmological constraint from the observed RHN DM relic density.
The (red) upper solid (dashed) line shows the XENON1T~\cite{Aprile:2017iyp}
   (LUX 2016~\cite{Akerib:2016vxi}) upper bound on $\sigma_{\rm SI}$,
   and the (red) dotted line shows the prospective reach for the upper bound on $\sigma_{\rm SI}$
   in the LZ DM experiment~\cite{Szydagis:2016few}.
Figs.~\ref{Fig:xH_sigma}\subref{Fig:xH_sigma_mzp=3750},
   \ref{Fig:xH_sigma}\subref{Fig:xH_sigma_mzp=3500}
   and \ref{Fig:xH_sigma}\subref{Fig:xH_sigma_mzp=3000}
   are the same as Fig.~\ref{Fig:xH_sigma}\subref{Fig:xH_sigma_mzp=4000},
   but for $m_{Z^\prime}=3.75$ TeV, 3.5 TeV and 3 TeV
   corresponding to Fig.~\ref{Fig:xH_alpha}\subref{Fig:xH_alpha_mzp=3750},
   \ref{Fig:xH_alpha}\subref{Fig:xH_alpha_mzp=3500}
   and \ref{Fig:xH_alpha}\subref{Fig:xH_alpha_mzp=3000}, respectively.
Fig.~\ref{Fig:xH_sigma}\subref{Fig:xH_sigma_mzp=3750} has
   a (green) shaded region in the range of $-0.9 \lesssim x_H \lesssim -0.5$
   to satisfy the three constraints,
   while Figs.~\ref{Fig:xH_sigma}\subref{Fig:xH_sigma_mzp=3500}
   and \ref{Fig:xH_sigma}\subref{Fig:xH_sigma_mzp=3000}
   have no such region.

\section{Conclusions}
\label{Sec_conclusion}
We have considered the DM scenario in the context
   of the classically conformal U(1)$^\prime$ extended SM,
   with three RHNs and the U(1)$^\prime$ Higgs field.
The model is free from all the U(1)$^\prime$ gauge and gravitational anomalies
   in the presence of the three RHNs.
We have introduced a $Z_2$-parity in the model,
   under which an odd-parity is assigned to one RHN,
   while all the other particles are assigned to be $Z_2$-even.
In our model, the $Z_2$-odd RHN serves as a stable DM candidate, 
   while the other two RHNs are utilized for the the minimal seesaw mechanism
   in order to reproduce the neutrino oscillation data
   and the observed baryon asymmetry in the Universe through leptogenesis. 
In this model, the U(1)$^\prime$ gauge symmetry is radiatively broken
   through the CW mechanism,
   by which the electroweak symmetry breaking is triggered.
There are three free parameters in our model,
   the U(1)$^\prime$ charge of the SM Higgs doublet ($x_H$),
   the new U(1)$^\prime$ gauge coupling ($\alpha_{g_X}$),
   and the U(1)$^\prime$ gauge boson ($Z^\prime$) mass ($m_{Z^\prime}$).

In this model context, 
   we have first investigated a possibility to resolve the electroweak vacuum instability
   with the current world average of the experimental data,
   $m_t =173.34$ GeV and $m_h=125.09$ GeV. 
By analyzing the RG evolutions
   of the couplings of the model at the two-loop level,
   we have performed a parameter scan 
   for the three parameters, $m_{Z^\prime}$, $\alpha_{g_X}$ and $x_H$,
   and have identified parameter regions which can solve the electroweak instability problem 
   and keep all coupling values in the perturbative regime up to the Planck mass scale.  
We have found that the resultant parameter regions are very severely constrained.  
Next, we have calculated the thermal relic density of the RHN DM
  and identified the model parameter region to reproduce the observed DM relic density
  of the Planck 2015 measurement.
In our model, the RHN DM particles mainly annihilate into the SM particles 
   through the $s$-channel process mediated by the $Z^\prime$ boson.
We have obtained the lower bound
   on $\alpha_{g_X}$ as a function of $m_{Z^\prime}$ and $x_H$ from the observed DM relic density. 
We have also considered the LHC Run-2 bounds
   from the search for the $Z^\prime$ boson resonance
   by the recent ATLAS and CMS analysis,
   which lead to the upper bounds on $\alpha_{g_X}$ as a function of $m_{Z^\prime}$ and $x_H$.
The LEP results from the search for effective 4-Fermi interactions mediated by the $Z^\prime$ boson
   can also constrain the model parameter space, but the LEP constraints are found to be weaker than 
   those obtained from the LHC Run-2 results.  
Finally, we have combined all the constraints.
The cosmological constraint on the RHN DM yields the lower bound on $\alpha_{g_X}$ as a function of $m_{Z^\prime}$ and $x_H$,
   while the upper bound on $\alpha_{g_X}$ is obtained from the LHC Run-2 results,
   so that these constraints are complementary to narrow the allowed parameter regions.
We have found that only small portions in these allowed parameter regions
   can solve the electroweak vacuum instability problem.
In particular, no allowed region to satisfy all constraints exists for $m_{Z^\prime} \lesssim 3.5$ TeV.
For the obtained allowed regions,
   we have calculated the spin-independent cross section of the RHN DM with nucleons.
We have found that the resultant cross section well below the current experimental upper bounds.

\section*{Acknowledgements}
The work of D.-s.T. and S.O. was supported by Advanced Medical Instrumentation unit [Sugawara unit]
  and Mathematical and Theoretical Physics unit [Hikami unit], respectively,
  of the Okinawa Institute of Science and Technology Graduate University.
The work of N.O. was supported in part by the United States Department of Energy (DE-SC0013680).


\bibliographystyle{utphys} 

\bibliography{../../U(1)'}                           

\providecommand{\href}[2]{#2}\begingroup\raggedright\begin{thebibliography}{100}

\bibitem{Minkowski:1977sc}
P.~Minkowski, ``{$\mu \to e\gamma$ at a Rate of One Out of $10^{9}$ Muon
  Decays?},''
\href{http://dx.doi.org/10.1016/0370-2693(77)90435-X}{Phys. Lett. {\bfseries
  B67} (1977) 421--428}.

\bibitem{Yanagida:1979as}
T.~Yanagida, ``{HORIZONTAL SYMMETRY AND MASSES OF NEUTRINOS},''
Conf. Proc. {\bfseries C7902131} (1979) 95--99.

\bibitem{GellMann:1980vs}
M.~Gell-Mann, P.~Ramond, and R.~Slansky, ``{Complex Spinors and Unified
  Theories},'' Conf. Proc. {\bfseries C790927} (1979) 315--321,
\href{http://arxiv.org/abs/1306.4669}{{\ttfamily arXiv:1306.4669 [hep-th]}}.

\bibitem{Glashow:1979nm}
S.~L. Glashow, ``{The Future of Elementary Particle Physics},''
\href{http://dx.doi.org/10.1007/978-1-4684-7197-7_15}{NATO Sci. Ser. B
  {\bfseries 61} (1980) 687}.

\bibitem{Mohapatra:1979ia}
R.~N. Mohapatra and G.~Senjanovic, ``{Neutrino Mass and Spontaneous Parity
  Violation},''
\href{http://dx.doi.org/10.1103/PhysRevLett.44.912}{Phys. Rev. Lett. {\bfseries
  44} (1980) 912}.

\bibitem{Mohapatra:1980qe}
R.~N. Mohapatra and R.~E. Marshak, ``{Local B-L Symmetry of Electroweak
  Interactions, Majorana Neutrinos and Neutron Oscillations},''
  \href{http://dx.doi.org/10.1103/PhysRevLett.44.1316}{Phys. Rev. Lett.
  {\bfseries 44} (1980) 1316--1319}.
[Erratum: Phys. Rev. Lett.44,1643(1980)].

\bibitem{Marshak:1979fm}
R.~E. Marshak and R.~N. Mohapatra, ``{Quark - Lepton Symmetry and B-L as the
  U(1) Generator of the Electroweak Symmetry Group},''
\href{http://dx.doi.org/10.1016/0370-2693(80)90436-0}{Phys. Lett. {\bfseries
  B91} (1980) 222--224}.

\bibitem{Wetterich:1981bx}
C.~Wetterich, ``{Neutrino Masses and the Scale of B-L Violation},''
\href{http://dx.doi.org/10.1016/0550-3213(81)90279-0}{Nucl. Phys. {\bfseries
  B187} (1981) 343--375}.

\bibitem{Masiero:1982fi}
A.~Masiero, J.~F. Nieves, and T.~Yanagida, ``{$B-L$ Violating Proton Decay and
  Late Cosmological Baryon Production},''
\href{http://dx.doi.org/10.1016/0370-2693(82)90024-7}{Phys. Lett. {\bfseries
  B116} (1982) 11--15}.

\bibitem{Mohapatra:1982xz}
R.~N. Mohapatra and G.~Senjanovic, ``{Spontaneous Breaking of Global $B-L$
  Symmetry and Matter - Antimatter Oscillations in Grand Unified Theories},''
\href{http://dx.doi.org/10.1103/PhysRevD.27.254}{Phys. Rev. {\bfseries D27}
  (1983) 254}.

\bibitem{Buchmuller:1991ce}
W.~Buchmuller, C.~Greub, and P.~Minkowski, ``{Neutrino masses, neutral vector
  bosons and the scale of B-L breaking},''
\href{http://dx.doi.org/10.1016/0370-2693(91)90952-M}{Phys. Lett. {\bfseries
  B267} (1991) 395--399}.

\bibitem{Appelquist:2002mw}
T.~Appelquist, B.~A. Dobrescu, and A.~R. Hopper, ``{Nonexotic neutral gauge
  bosons},'' \href{http://dx.doi.org/10.1103/PhysRevD.68.035012}{Phys. Rev.
  {\bfseries D68} (2003) 035012},
\href{http://arxiv.org/abs/hep-ph/0212073}{{\ttfamily arXiv:hep-ph/0212073
  [hep-ph]}}.

\bibitem{Oda:2015gna}
S.~Oda, N.~Okada, and D.-s. Takahashi, ``{Classically conformal U(1)$^\prime$
  extended standard model and Higgs vacuum stability},''
  \href{http://dx.doi.org/10.1103/PhysRevD.92.015026}{Phys. Rev. {\bfseries
  D92} no.~1, (2015) 015026},
\href{http://arxiv.org/abs/1504.06291}{{\ttfamily arXiv:1504.06291 [hep-ph]}}.

\bibitem{Das:2016zue}
A.~Das, S.~Oda, N.~Okada, and D.-s. Takahashi, ``{Classically conformal
  U(1)$^\prime$ extended standard model, electroweak vacuum stability, and LHC
  Run-2 bounds},'' \href{http://dx.doi.org/10.1103/PhysRevD.93.115038}{Phys.
  Rev. {\bfseries D93} no.~11, (2016) 115038},
\href{http://arxiv.org/abs/1605.01157}{{\ttfamily arXiv:1605.01157 [hep-ph]}}.

\bibitem{Hempfling:1996ht}
R.~Hempfling, ``{The Next-to-minimal Coleman-Weinberg model},''
  \href{http://dx.doi.org/10.1016/0370-2693(96)00446-7}{Phys. Lett. {\bfseries
  B379} (1996) 153--158},
\href{http://arxiv.org/abs/hep-ph/9604278}{{\ttfamily arXiv:hep-ph/9604278
  [hep-ph]}}.

\bibitem{Dias:2006th}
A.~G. Dias, ``{Neutrino Mass Through Concomitant Breakdown of the U(1) Chiral
  and Scale Symmetries},''
  \href{http://dx.doi.org/10.1103/PhysRevD.73.096002}{Phys. Rev. {\bfseries
  D73} (2006) 096002},
\href{http://arxiv.org/abs/hep-ph/0604219}{{\ttfamily arXiv:hep-ph/0604219
  [hep-ph]}}.

\bibitem{Espinosa:2007qk}
J.~R. Espinosa and M.~Quiros, ``{Novel Effects in Electroweak Breaking from a
  Hidden Sector},'' \href{http://dx.doi.org/10.1103/PhysRevD.76.076004}{Phys.
  Rev. {\bfseries D76} (2007) 076004},
\href{http://arxiv.org/abs/hep-ph/0701145}{{\ttfamily arXiv:hep-ph/0701145
  [hep-ph]}}.

\bibitem{Chang:2007ki}
W.-F. Chang, J.~N. Ng, and J.~M.~S. Wu, ``{Shadow Higgs from a scale-invariant
  hidden U(1)(s) model},''
  \href{http://dx.doi.org/10.1103/PhysRevD.75.115016}{Phys. Rev. {\bfseries
  D75} (2007) 115016},
\href{http://arxiv.org/abs/hep-ph/0701254}{{\ttfamily arXiv:hep-ph/0701254
  [HEP-PH]}}.

\bibitem{Foot:2007as}
R.~Foot, A.~Kobakhidze, and R.~R. Volkas, ``{Electroweak Higgs as a
  pseudo-Goldstone boson of broken scale invariance},''
  \href{http://dx.doi.org/10.1016/j.physletb.2007.06.084}{Phys. Lett.
  {\bfseries B655} (2007) 156--161},
\href{http://arxiv.org/abs/0704.1165}{{\ttfamily arXiv:0704.1165 [hep-ph]}}.

\bibitem{Foot:2007ay}
R.~Foot, A.~Kobakhidze, K.~McDonald, and R.~Volkas, ``{Neutrino mass in
  radiatively-broken scale-invariant models},''
  \href{http://dx.doi.org/10.1103/PhysRevD.76.075014}{Phys. Rev. {\bfseries
  D76} (2007) 075014},
\href{http://arxiv.org/abs/0706.1829}{{\ttfamily arXiv:0706.1829 [hep-ph]}}.

\bibitem{Meissner:2006zh}
K.~A. Meissner and H.~Nicolai, ``{Conformal Symmetry and the Standard Model},''
  \href{http://dx.doi.org/10.1016/j.physletb.2007.03.023}{Phys. Lett.
  {\bfseries B648} (2007) 312--317},
\href{http://arxiv.org/abs/hep-th/0612165}{{\ttfamily arXiv:hep-th/0612165
  [hep-th]}}.

\bibitem{Foot:2007iy}
R.~Foot, A.~Kobakhidze, K.~L. McDonald, and R.~R. Volkas, ``{A Solution to the
  hierarchy problem from an almost decoupled hidden sector within a classically
  scale invariant theory},''
  \href{http://dx.doi.org/10.1103/PhysRevD.77.035006}{Phys. Rev. {\bfseries
  D77} (2008) 035006},
\href{http://arxiv.org/abs/0709.2750}{{\ttfamily arXiv:0709.2750 [hep-ph]}}.

\bibitem{Meissner:2007xv}
K.~A. Meissner and H.~Nicolai, ``{Effective action, conformal anomaly and the
  issue of quadratic divergences},''
  \href{http://dx.doi.org/10.1016/j.physletb.2007.12.035}{Phys. Lett.
  {\bfseries B660} (2008) 260--266},
\href{http://arxiv.org/abs/0710.2840}{{\ttfamily arXiv:0710.2840 [hep-th]}}.

\bibitem{Meissner:2008gj}
K.~A. Meissner and H.~Nicolai, ``{Neutrinos, Axions and Conformal Symmetry},''
  \href{http://dx.doi.org/10.1140/epjc/s10052-008-0760-x}{Eur. Phys. J.
  {\bfseries C57} (2008) 493--498},
\href{http://arxiv.org/abs/0803.2814}{{\ttfamily arXiv:0803.2814 [hep-th]}}.

\bibitem{Iso:2009ss}
S.~Iso, N.~Okada, and Y.~Orikasa, ``{Classically conformal $B-L$ extended
  Standard Model},''
  \href{http://dx.doi.org/10.1016/j.physletb.2009.04.046}{Phys. Lett.
  {\bfseries B676} (2009) 81--87},
\href{http://arxiv.org/abs/0902.4050}{{\ttfamily arXiv:0902.4050 [hep-ph]}}.

\bibitem{Iso:2009nw}
S.~Iso, N.~Okada, and Y.~Orikasa, ``{The minimal B-L model naturally realized
  at TeV scale},'' \href{http://dx.doi.org/10.1103/PhysRevD.80.115007}{Phys.
  Rev. {\bfseries D80} (2009) 115007},
\href{http://arxiv.org/abs/0909.0128}{{\ttfamily arXiv:0909.0128 [hep-ph]}}.

\bibitem{Holthausen:2009uc}
M.~Holthausen, M.~Lindner, and M.~A. Schmidt, ``{Radiative Symmetry Breaking of
  the Minimal Left-Right Symmetric Model},''
  \href{http://dx.doi.org/10.1103/PhysRevD.82.055002}{Phys. Rev. {\bfseries
  D82} (2010) 055002},
\href{http://arxiv.org/abs/0911.0710}{{\ttfamily arXiv:0911.0710 [hep-ph]}}.

\bibitem{Farzinnia:2013pga}
A.~Farzinnia, H.-J. He, and J.~Ren, ``{Natural Electroweak Symmetry Breaking
  from Scale Invariant Higgs Mechanism},''
  \href{http://dx.doi.org/10.1016/j.physletb.2013.09.060}{Phys. Lett.
  {\bfseries B727} (2013) 141--150},
\href{http://arxiv.org/abs/1308.0295}{{\ttfamily arXiv:1308.0295 [hep-ph]}}.

\bibitem{Heikinheimo:2013fta}
M.~Heikinheimo, A.~Racioppi, M.~Raidal, C.~Spethmann, and K.~Tuominen,
  ``{Physical Naturalness and Dynamical Breaking of Classical Scale
  Invariance},'' \href{http://dx.doi.org/10.1142/S0217732314500771}{Mod. Phys.
  Lett. {\bfseries A29} (2014) 1450077},
\href{http://arxiv.org/abs/1304.7006}{{\ttfamily arXiv:1304.7006 [hep-ph]}}.

\bibitem{Farzinnia:2014xia}
A.~Farzinnia and J.~Ren, ``{Higgs Partner Searches and Dark Matter
  Phenomenology in a Classically Scale Invariant Higgs Boson Sector},''
  \href{http://dx.doi.org/10.1103/PhysRevD.90.015019}{Phys. Rev. {\bfseries
  D90} no.~1, (2014) 015019},
\href{http://arxiv.org/abs/1405.0498}{{\ttfamily arXiv:1405.0498 [hep-ph]}}.

\bibitem{Lindner:2014oea}
M.~Lindner, S.~Schmidt, and J.~Smirnov, ``{Neutrino Masses and Conformal
  Electro-Weak Symmetry Breaking},''
  \href{http://dx.doi.org/10.1007/JHEP10(2014)177}{JHEP {\bfseries 10} (2014)
  177},
\href{http://arxiv.org/abs/1405.6204}{{\ttfamily arXiv:1405.6204 [hep-ph]}}.

\bibitem{Khoze:2014xha}
V.~V. Khoze, C.~McCabe, and G.~Ro, ``{Higgs vacuum stability from the dark
  matter portal},'' \href{http://dx.doi.org/10.1007/JHEP08(2014)026}{JHEP
  {\bfseries 08} (2014) 026},
\href{http://arxiv.org/abs/1403.4953}{{\ttfamily arXiv:1403.4953 [hep-ph]}}.

\bibitem{Gabrielli:2013hma}
E.~Gabrielli, M.~Heikinheimo, K.~Kannike, A.~Racioppi, M.~Raidal, and
  C.~Spethmann, ``{Towards Completing the Standard Model: Vacuum Stability,
  EWSB and Dark Matter},''
  \href{http://dx.doi.org/10.1103/PhysRevD.89.015017}{Phys. Rev. {\bfseries
  D89} no.~1, (2014) 015017},
\href{http://arxiv.org/abs/1309.6632}{{\ttfamily arXiv:1309.6632 [hep-ph]}}.

\bibitem{Altmannshofer:2014vra}
W.~Altmannshofer, W.~A. Bardeen, M.~Bauer, M.~Carena, and J.~D. Lykken,
  ``{Light Dark Matter, Naturalness, and the Radiative Origin of the
  Electroweak Scale},'' \href{http://dx.doi.org/10.1007/JHEP01(2015)032}{JHEP
  {\bfseries 01} (2015) 032},
\href{http://arxiv.org/abs/1408.3429}{{\ttfamily arXiv:1408.3429 [hep-ph]}}.

\bibitem{Karam:2015jta}
A.~Karam and K.~Tamvakis, ``{Dark matter and neutrino masses from a
  scale-invariant multi-Higgs portal},''
  \href{http://dx.doi.org/10.1103/PhysRevD.92.075010}{Phys. Rev. {\bfseries
  D92} no.~7, (2015) 075010},
\href{http://arxiv.org/abs/1508.03031}{{\ttfamily arXiv:1508.03031 [hep-ph]}}.

\bibitem{Haba:2015yfa}
N.~Haba, H.~Ishida, N.~Okada, and Y.~Yamaguchi, ``{Electroweak symmetry
  breaking through bosonic seesaw mechanism in a classically conformal
  extension of the Standard Model},''
\href{http://arxiv.org/abs/1509.01923}{{\ttfamily arXiv:1509.01923 [hep-ph]}}.

\bibitem{Okada:2015gia}
H.~Okada, Y.~Orikasa, and K.~Yagyu, ``{Higgs Triplet Model with Classically
  Conformal Invariance},''
\href{http://arxiv.org/abs/1510.00799}{{\ttfamily arXiv:1510.00799 [hep-ph]}}.

\bibitem{Latosinski:2015pba}
A.~Latosinski, A.~Lewandowski, K.~A. Meissner, and H.~Nicolai, ``{Conformal
  Standard Model with an extended scalar sector},''
  \href{http://dx.doi.org/10.1007/JHEP10(2015)170}{JHEP {\bfseries 10} (2015)
  170},
\href{http://arxiv.org/abs/1507.01755}{{\ttfamily arXiv:1507.01755 [hep-ph]}}.

\bibitem{Wang:2015sxe}
Z.-W. Wang, F.~S. Sage, T.~G. Steele, and R.~B. Mann, ``{Asymptotic Safety in
  the Conformal Hidden Sector?},''
\href{http://arxiv.org/abs/1511.02531}{{\ttfamily arXiv:1511.02531 [hep-ph]}}.

\bibitem{Goertz:2015dba}
F.~Goertz, ``{Electroweak Symmetry Breaking without the $\mu^2$ Term},''
  \href{http://dx.doi.org/10.1103/PhysRevD.94.015013}{Phys. Rev. {\bfseries
  D94} no.~1, (2016) 015013},
\href{http://arxiv.org/abs/1504.00355}{{\ttfamily arXiv:1504.00355 [hep-ph]}}.

\bibitem{Haba:2015lka}
N.~Haba, H.~Ishida, N.~Okada, and Y.~Yamaguchi, ``{Bosonic seesaw mechanism in
  a classically conformal extension of the Standard Model},''
  \href{http://dx.doi.org/10.1016/j.physletb.2016.01.050}{Phys. Lett.
  {\bfseries B754} (2016) 349--352},
\href{http://arxiv.org/abs/1508.06828}{{\ttfamily arXiv:1508.06828 [hep-ph]}}.

\bibitem{Haba:2015nwl}
N.~Haba, H.~Ishida, R.~Takahashi, and Y.~Yamaguchi, ``{Gauge coupling
  unification in a classically scale invariant model},''
  \href{http://dx.doi.org/10.1007/JHEP02(2016)058}{JHEP {\bfseries 02} (2016)
  058},
\href{http://arxiv.org/abs/1511.02107}{{\ttfamily arXiv:1511.02107 [hep-ph]}}.

\bibitem{Ghorbani:2015xvz}
K.~Ghorbani and H.~Ghorbani, ``{Scalar Dark Matter in Scale Invariant Standard
  Model},'' \href{http://dx.doi.org/10.1007/JHEP04(2016)024}{JHEP {\bfseries
  04} (2016) 024},
\href{http://arxiv.org/abs/1511.08432}{{\ttfamily arXiv:1511.08432 [hep-ph]}}.

\bibitem{Haba:2015qbz}
N.~Haba, H.~Ishida, N.~Kitazawa, and Y.~Yamaguchi, ``{A new dynamics of
  electroweak symmetry breaking with classically scale invariance},''
  \href{http://dx.doi.org/10.1016/j.physletb.2016.02.052}{Phys. Lett.
  {\bfseries B755} (2016) 439--443},
\href{http://arxiv.org/abs/1512.05061}{{\ttfamily arXiv:1512.05061 [hep-ph]}}.

\bibitem{Ahriche:2015loa}
A.~Ahriche, K.~L. McDonald, and S.~Nasri, ``{A Radiative Model for the Weak
  Scale and Neutrino Mass via Dark Matter},''
  \href{http://dx.doi.org/10.1007/JHEP02(2016)038}{JHEP {\bfseries 02} (2016)
  038},
\href{http://arxiv.org/abs/1508.02607}{{\ttfamily arXiv:1508.02607 [hep-ph]}}.

\bibitem{Ishida:2016ogu}
H.~Ishida, S.~Matsuzaki, and Y.~Yamaguchi, ``{Invisible axionlike dark matter
  from the electroweak bosonic seesaw mechanism},''
  \href{http://dx.doi.org/10.1103/PhysRevD.94.095011}{Phys. Rev. {\bfseries
  D94} no.~9, (2016) 095011},
\href{http://arxiv.org/abs/1604.07712}{{\ttfamily arXiv:1604.07712 [hep-ph]}}.

\bibitem{Hatanaka:2016rek}
H.~Hatanaka, D.-W. Jung, and P.~Ko, ``{AdS/QCD approach to the scale-invariant
  extension of the standard model with a strongly interacting hidden sector},''
  \href{http://dx.doi.org/10.1007/JHEP08(2016)094}{JHEP {\bfseries 08} (2016)
  094},
\href{http://arxiv.org/abs/1606.02969}{{\ttfamily arXiv:1606.02969 [hep-ph]}}.

\bibitem{Karam:2016rsz}
A.~Karam and K.~Tamvakis, ``{Dark Matter from a Classically Scale-Invariant
  $SU(3)_X$},'' \href{http://dx.doi.org/10.1103/PhysRevD.94.055004}{Phys. Rev.
  {\bfseries D94} no.~5, (2016) 055004},
\href{http://arxiv.org/abs/1607.01001}{{\ttfamily arXiv:1607.01001 [hep-ph]}}.

\bibitem{Marzola:2016xgb}
L.~Marzola and A.~Racioppi, ``{Minimal but non-minimal inflation and
  electroweak symmetry breaking},''
  \href{http://dx.doi.org/10.1088/1475-7516/2016/10/010}{JCAP {\bfseries 1610}
  no.~10, (2016) 010},
\href{http://arxiv.org/abs/1606.06887}{{\ttfamily arXiv:1606.06887 [hep-ph]}}.

\bibitem{Das:2015nwk}
A.~Das, N.~Okada, and N.~Papapietro, ``{Electroweak vacuum stability in
  classically conformal B-L extension of the Standard Model},''
  \href{http://dx.doi.org/10.1140/epjc/s10052-017-4683-2}{Eur. Phys. J.
  {\bfseries C77} no.~2, (2017) 122},
\href{http://arxiv.org/abs/1509.01466}{{\ttfamily arXiv:1509.01466 [hep-ph]}}.

\bibitem{Kannike:2016wuy}
K.~Kannike, M.~Raidal, C.~Spethmann, and H.~Veerm^^c3^^a4e, ``{Evolving Planck
  Mass in Classically Scale-Invariant Theories},''
  \href{http://dx.doi.org/10.1007/JHEP04(2017)026}{JHEP {\bfseries 04} (2017)
  026},
\href{http://arxiv.org/abs/1610.06571}{{\ttfamily arXiv:1610.06571 [hep-ph]}}.

\bibitem{Marzola:2017jzl}
L.~Marzola, A.~Racioppi, and V.~Vaskonen, ``{Phase transition and gravitational
  wave phenomenology of scalar conformal extensions of the Standard Model},''
  \href{http://dx.doi.org/10.1140/epjc/s10052-017-4996-1}{Eur. Phys. J.
  {\bfseries C77} no.~7, (2017) 484},
\href{http://arxiv.org/abs/1704.01034}{{\ttfamily arXiv:1704.01034 [hep-ph]}}.

\bibitem{Coleman:1973jx}
S.~R. Coleman and E.~J. Weinberg, ``{Radiative Corrections as the Origin of
  Spontaneous Symmetry Breaking},''
\href{http://dx.doi.org/10.1103/PhysRevD.7.1888}{Phys. Rev. {\bfseries D7}
  (1973) 1888--1910}.

\bibitem{ATLAS:2014wva}
{\bfseries ATLAS, CDF, CMS, D0} Collaboration, ``{First combination of Tevatron
  and LHC measurements of the top-quark mass},''
\href{http://arxiv.org/abs/1403.4427}{{\ttfamily arXiv:1403.4427 [hep-ex]}}.

\bibitem{Aad:2015zhl}
{\bfseries ATLAS, CMS} Collaboration, G.~Aad {\em et~al.}, ``{Combined
  Measurement of the Higgs Boson Mass in $pp$ Collisions at $\sqrt{s}=7$ and 8
  TeV with the ATLAS and CMS Experiments},''
  \href{http://dx.doi.org/10.1103/PhysRevLett.114.191803}{Phys. Rev. Lett.
  {\bfseries 114} (2015) 191803},
\href{http://arxiv.org/abs/1503.07589}{{\ttfamily arXiv:1503.07589 [hep-ex]}}.

\bibitem{TheATLAScollaboration:2015jgi}
T.~A. collaboration, ``{Search for new phenomena in the dilepton final state
  using proton-proton collisions at $\sqrt{s}$ = 13 TeV with the ATLAS
  detector},''
ATLAS-CONF-2015-070.

\bibitem{CMS:2015nhc}
{\bfseries CMS} Collaboration, C.~Collaboration, ``{Search for a Narrow
  Resonance Produced in 13 TeV pp Collisions Decaying to Electron Pair or Muon
  Pair Final States},''
CMS-PAS-EXO-15-005.

\bibitem{Okada:2010wd}
N.~Okada and O.~Seto, ``{Higgs portal dark matter in the minimal gauged
  $U(1)_{B-L}$ model},''
  \href{http://dx.doi.org/10.1103/PhysRevD.82.023507}{Phys. Rev. {\bfseries
  D82} (2010) 023507},
\href{http://arxiv.org/abs/1002.2525}{{\ttfamily arXiv:1002.2525 [hep-ph]}}.

\bibitem{Anisimov:2008gg}
A.~Anisimov and P.~Di~Bari, ``{Cold Dark Matter from heavy Right-Handed
  neutrino mixing},'' \href{http://dx.doi.org/10.1103/PhysRevD.80.073017}{Phys.
  Rev. {\bfseries D80} (2009) 073017},
\href{http://arxiv.org/abs/0812.5085}{{\ttfamily arXiv:0812.5085 [hep-ph]}}.

\bibitem{Fukugita:1986hr}
M.~Fukugita and T.~Yanagida, ``{Baryogenesis Without Grand Unification},''
\href{http://dx.doi.org/10.1016/0370-2693(86)91126-3}{Phys. Lett. {\bfseries
  B174} (1986) 45--47}.

\bibitem{King:1999mb}
S.~F. King, ``{Large mixing angle MSW and atmospheric neutrinos from single
  right-handed neutrino dominance and U(1) family symmetry},''
  \href{http://dx.doi.org/10.1016/S0550-3213(00)00109-7}{Nucl. Phys. {\bfseries
  B576} (2000) 85--105},
\href{http://arxiv.org/abs/hep-ph/9912492}{{\ttfamily arXiv:hep-ph/9912492
  [hep-ph]}}.

\bibitem{Frampton:2002qc}
P.~H. Frampton, S.~L. Glashow, and T.~Yanagida, ``{Cosmological sign of
  neutrino CP violation},''
  \href{http://dx.doi.org/10.1016/S0370-2693(02)02853-8}{Phys. Lett. {\bfseries
  B548} (2002) 119--121},
\href{http://arxiv.org/abs/hep-ph/0208157}{{\ttfamily arXiv:hep-ph/0208157
  [hep-ph]}}.

\bibitem{Burell:2011wh}
Z.~M. Burell and N.~Okada, ``{Supersymmetric minimal B-L model at the TeV scale
  with right-handed Majorana neutrino dark matter},''
  \href{http://dx.doi.org/10.1103/PhysRevD.85.055011}{Phys. Rev. {\bfseries
  D85} (2012) 055011},
\href{http://arxiv.org/abs/1111.1789}{{\ttfamily arXiv:1111.1789 [hep-ph]}}.

\bibitem{Basso:2012ti}
L.~Basso, O.~Fischer, and J.~J. van~der Bij, ``{Natural $Z^\prime$ model with
  an inverse seesaw mechanism and leptonic dark matter},''
  \href{http://dx.doi.org/10.1103/PhysRevD.87.035015}{Phys. Rev. {\bfseries
  D87} no.~3, (2013) 035015},
\href{http://arxiv.org/abs/1207.3250}{{\ttfamily arXiv:1207.3250 [hep-ph]}}.

\bibitem{Dudas:2013sia}
E.~Dudas, L.~Heurtier, Y.~Mambrini, and B.~Zaldivar, ``{Extra U(1), effective
  operators, anomalies and dark matter},''
  \href{http://dx.doi.org/10.1007/JHEP11(2013)083}{JHEP {\bfseries 11} (2013)
  083},
\href{http://arxiv.org/abs/1307.0005}{{\ttfamily arXiv:1307.0005 [hep-ph]}}.

\bibitem{Das:2013jca}
M.~Das and S.~Mohanty, ``{Leptophilic dark matter in gauged $L_{\mu}-L_{\tau}$
  extension of MSSM},''
  \href{http://dx.doi.org/10.1103/PhysRevD.89.025004}{Phys. Rev. {\bfseries
  D89} no.~2, (2014) 025004},
\href{http://arxiv.org/abs/1306.4505}{{\ttfamily arXiv:1306.4505 [hep-ph]}}.

\bibitem{Chu:2013jja}
X.~Chu, Y.~Mambrini, J.~Quevillon, and B.~Zaldivar, ``{Thermal and non-thermal
  production of dark matter via $Z^\prime$-portal(s)},''
  \href{http://dx.doi.org/10.1088/1475-7516/2014/01/034}{JCAP {\bfseries 1401}
  (2014) 034},
\href{http://arxiv.org/abs/1306.4677}{{\ttfamily arXiv:1306.4677 [hep-ph]}}.

\bibitem{Lindner:2013awa}
M.~Lindner, D.~Schmidt, and A.~Watanabe, ``{Dark matter and $U(1)'$ symmetry
  for the right-handed neutrinos},''
  \href{http://dx.doi.org/10.1103/PhysRevD.89.013007}{Phys. Rev. {\bfseries
  D89} no.~1, (2014) 013007},
\href{http://arxiv.org/abs/1310.6582}{{\ttfamily arXiv:1310.6582 [hep-ph]}}.

\bibitem{Alves:2013tqa}
A.~Alves, S.~Profumo, and F.~S. Queiroz, ``{The dark $Z^\prime$ portal: direct,
  indirect and collider searches},''
  \href{http://dx.doi.org/10.1007/JHEP04(2014)063}{JHEP {\bfseries 04} (2014)
  063},
\href{http://arxiv.org/abs/1312.5281}{{\ttfamily arXiv:1312.5281 [hep-ph]}}.

\bibitem{Kopp:2014tsa}
J.~Kopp, L.~Michaels, and J.~Smirnov, ``{Loopy Constraints on Leptophilic Dark
  Matter and Internal Bremsstrahlung},''
  \href{http://dx.doi.org/10.1088/1475-7516/2014/04/022}{JCAP {\bfseries 1404}
  (2014) 022},
\href{http://arxiv.org/abs/1401.6457}{{\ttfamily arXiv:1401.6457 [hep-ph]}}.

\bibitem{Agrawal:2014ufa}
P.~Agrawal, Z.~Chacko, and C.~B. Verhaaren, ``{Leptophilic Dark Matter and the
  Anomalous Magnetic Moment of the Muon},''
  \href{http://dx.doi.org/10.1007/JHEP08(2014)147}{JHEP {\bfseries 08} (2014)
  147},
\href{http://arxiv.org/abs/1402.7369}{{\ttfamily arXiv:1402.7369 [hep-ph]}}.

\bibitem{Hooper:2014fda}
D.~Hooper, ``{$Z^\prime$ Mediated Dark Matter Models for the Galactic Center
  Gamma-Ray Excess},''
  \href{http://dx.doi.org/10.1103/PhysRevD.91.035025}{Phys. Rev. {\bfseries
  D91} (2015) 035025},
\href{http://arxiv.org/abs/1411.4079}{{\ttfamily arXiv:1411.4079 [hep-ph]}}.

\bibitem{Ma:2014qra}
E.~Ma and R.~Srivastava, ``{Dirac or inverse seesaw neutrino masses with $B-L$
  gauge symmetry and $S_3$ flavor symmetry},''
  \href{http://dx.doi.org/10.1016/j.physletb.2014.12.049}{Phys. Lett.
  {\bfseries B741} (2015) 217--222},
\href{http://arxiv.org/abs/1411.5042}{{\ttfamily arXiv:1411.5042 [hep-ph]}}.

\bibitem{Alves:2015pea}
A.~Alves, A.~Berlin, S.~Profumo, and F.~S. Queiroz, ``{Dark Matter
  Complementarity and the $Z^\prime$ Portal},''
  \href{http://dx.doi.org/10.1103/PhysRevD.92.083004}{Phys. Rev. {\bfseries
  D92} no.~8, (2015) 083004},
\href{http://arxiv.org/abs/1501.03490}{{\ttfamily arXiv:1501.03490 [hep-ph]}}.

\bibitem{Ghorbani:2015baa}
K.~Ghorbani and H.~Ghorbani, ``{Two-portal Dark Matter},''
  \href{http://dx.doi.org/10.1103/PhysRevD.91.123541}{Phys. Rev. {\bfseries
  D91} no.~12, (2015) 123541},
\href{http://arxiv.org/abs/1504.03610}{{\ttfamily arXiv:1504.03610 [hep-ph]}}.

\bibitem{Sanchez-Vega:2015qva}
B.~L. S^^c3^^a1nchez-Vega and E.~R. Schmitz, ``{Fermionic dark matter and
  neutrino masses in a B-L model},''
  \href{http://dx.doi.org/10.1103/PhysRevD.92.053007}{Phys. Rev. {\bfseries
  D92} (2015) 053007},
\href{http://arxiv.org/abs/1505.03595}{{\ttfamily arXiv:1505.03595 [hep-ph]}}.

\bibitem{Duerr:2015wfa}
M.~Duerr, P.~Fileviez~Perez, and J.~Smirnov, ``{Simplified Dirac Dark Matter
  Models and Gamma-Ray Lines},''
  \href{http://dx.doi.org/10.1103/PhysRevD.92.083521}{Phys. Rev. {\bfseries
  D92} no.~8, (2015) 083521},
\href{http://arxiv.org/abs/1506.05107}{{\ttfamily arXiv:1506.05107 [hep-ph]}}.

\bibitem{Alves:2015mua}
A.~Alves, A.~Berlin, S.~Profumo, and F.~S. Queiroz, ``{Dirac-fermionic dark
  matter in U(1)$_{X}$ models},''
  \href{http://dx.doi.org/10.1007/JHEP10(2015)076}{JHEP {\bfseries 10} (2015)
  076},
\href{http://arxiv.org/abs/1506.06767}{{\ttfamily arXiv:1506.06767 [hep-ph]}}.

\bibitem{Ma:2015mjd}
E.~Ma, N.~Pollard, R.~Srivastava, and M.~Zakeri, ``{Gauge $B-L$ Model with
  Residual $Z_3$ Symmetry},''
  \href{http://dx.doi.org/10.1016/j.physletb.2015.09.010}{Phys. Lett.
  {\bfseries B750} (2015) 135--138},
\href{http://arxiv.org/abs/1507.03943}{{\ttfamily arXiv:1507.03943 [hep-ph]}}.

\bibitem{Okada:2016gsh}
N.~Okada and S.~Okada, ``{$Z^\prime_{BL}$ portal dark matter and LHC Run-2
  results},'' \href{http://dx.doi.org/10.1103/PhysRevD.93.075003}{Phys. Rev.
  {\bfseries D93} no.~7, (2016) 075003},
\href{http://arxiv.org/abs/1601.07526}{{\ttfamily arXiv:1601.07526 [hep-ph]}}.

\bibitem{Okada:2016tzi}
N.~Okada and N.~Papapietro, ``{R-parity Conserving Minimal SUSY $B-L$ Model},''
\href{http://arxiv.org/abs/1603.01769}{{\ttfamily arXiv:1603.01769 [hep-ph]}}.

\bibitem{Chao:2016avy}
W.~Chao, H.-k. Guo, and Y.~Zhang, ``{Majorana Dark matter with B+L gauge
  symmetry},''
\href{http://arxiv.org/abs/1604.01771}{{\ttfamily arXiv:1604.01771 [hep-ph]}}.

\bibitem{Biswas:2016ewm}
A.~Biswas, S.~Choubey, and S.~Khan, ``{Galactic gamma ray excess and dark
  matter phenomenology in a $U(1)_{B-L}$ model},''
  \href{http://dx.doi.org/10.1007/JHEP08(2016)114}{JHEP {\bfseries 08} (2016)
  114},
\href{http://arxiv.org/abs/1604.06566}{{\ttfamily arXiv:1604.06566 [hep-ph]}}.

\bibitem{Accomando:2016sge}
E.~Accomando, C.~Coriano, L.~Delle~Rose, J.~Fiaschi, C.~Marzo, and S.~Moretti,
  ``{$Z^\prime$, Higgses and heavy neutrinos in U(1)$^\prime$ models: from the
  LHC to the GUT scale},''
  \href{http://dx.doi.org/10.1007/JHEP07(2016)086}{JHEP {\bfseries 07} (2016)
  086},
\href{http://arxiv.org/abs/1605.02910}{{\ttfamily arXiv:1605.02910 [hep-ph]}}.

\bibitem{Fairbairn:2016iuf}
M.~Fairbairn, J.~Heal, F.~Kahlhoefer, and P.~Tunney, ``{Constraints on
  $Z^\prime$ models from LHC dijet searches and implications for dark
  matter},'' \href{http://dx.doi.org/10.1007/JHEP09(2016)018}{JHEP {\bfseries
  09} (2016) 018},
\href{http://arxiv.org/abs/1605.07940}{{\ttfamily arXiv:1605.07940 [hep-ph]}}.

\bibitem{Klasen:2016qux}
M.~Klasen, F.~Lyonnet, and F.~S. Queiroz, ``{NLO+NLL Collider Bounds, Dirac
  Fermion and Scalar Dark Matter in the B-L Model},''
\href{http://arxiv.org/abs/1607.06468}{{\ttfamily arXiv:1607.06468 [hep-ph]}}.

\bibitem{Dev:2016xcp}
P.~S. Bhupal~Dev, R.~N. Mohapatra, and Y.~Zhang, ``{Naturally stable
  right-handed neutrino dark matter},''
  \href{http://dx.doi.org/10.1007/JHEP11(2016)077}{JHEP {\bfseries 11} (2016)
  077},
\href{http://arxiv.org/abs/1608.06266}{{\ttfamily arXiv:1608.06266 [hep-ph]}}.

\bibitem{Altmannshofer:2016jzy}
W.~Altmannshofer, S.~Gori, S.~Profumo, and F.~S. Queiroz, ``{Explaining dark
  matter and $B$ decay anomalies with an $L_\mu - L_\tau$ model},''
  \href{http://dx.doi.org/10.1007/JHEP12(2016)106}{JHEP {\bfseries 12} (2016)
  106},
\href{http://arxiv.org/abs/1609.04026}{{\ttfamily arXiv:1609.04026 [hep-ph]}}.

\bibitem{Okada:2016tci}
N.~Okada and S.~Okada, ``{$Z^\prime$-portal right-handed neutrino dark matter
  in the minimal U(1)$_X$ extended Standard Model},''
  \href{http://dx.doi.org/10.1103/PhysRevD.95.035025}{Phys. Rev. {\bfseries
  D95} no.~3, (2017) 035025},
\href{http://arxiv.org/abs/1611.02672}{{\ttfamily arXiv:1611.02672 [hep-ph]}}.

\bibitem{Kaneta:2016vkq}
K.~Kaneta, Z.~Kang, and H.-S. Lee, ``{Right-handed neutrino dark matter under
  the $B-L$ gauge interaction},''
  \href{http://dx.doi.org/10.1007/JHEP02(2017)031}{JHEP {\bfseries 02} (2017)
  031},
\href{http://arxiv.org/abs/1606.09317}{{\ttfamily arXiv:1606.09317 [hep-ph]}}.

\bibitem{Arcadi:2017kky}
G.~Arcadi, M.~Dutra, P.~Ghosh, M.~Lindner, Y.~Mambrini, M.~Pierre, S.~Profumo,
  and F.~S. Queiroz, ``{The Waning of the WIMP? A Review of Models, Searches,
  and Constraints},''
\href{http://arxiv.org/abs/1703.07364}{{\ttfamily arXiv:1703.07364 [hep-ph]}}.

\bibitem{Okada:2012sg}
N.~Okada and Y.~Orikasa, ``{Dark matter in the classically conformal B-L
  model},'' \href{http://dx.doi.org/10.1103/PhysRevD.85.115006}{Phys. Rev.
  {\bfseries D85} (2012) 115006},
\href{http://arxiv.org/abs/1202.1405}{{\ttfamily arXiv:1202.1405 [hep-ph]}}.

\bibitem{Basak:2013cga}
T.~Basak and T.~Mondal, ``{Constraining Minimal $U(1)_{B-L}$ model from Dark
  Matter Observations},''
  \href{http://dx.doi.org/10.1103/PhysRevD.89.063527}{Phys. Rev. {\bfseries
  D89} (2014) 063527},
\href{http://arxiv.org/abs/1308.0023}{{\ttfamily arXiv:1308.0023 [hep-ph]}}.

\bibitem{Aghanim:2015xee}
{\bfseries Planck} Collaboration, N.~Aghanim {\em et~al.}, ``{Planck 2015
  results. XI. CMB power spectra, likelihoods, and robustness of parameters},''
  \href{http://dx.doi.org/10.1051/0004-6361/201526926}{Astron. Astrophys.
  {\bfseries 594} (2016) A11},
\href{http://arxiv.org/abs/1507.02704}{{\ttfamily arXiv:1507.02704
  [astro-ph.CO]}}.

\bibitem{ATLAS:2016cyf}
{\bfseries ATLAS} Collaboration, T.~A. collaboration, ``{Search for new
  high-mass resonances in the dilepton final state using proton-proton
  collisions at $\sqrt{s}$ = 13 TeV with the ATLAS detector},''
ATLAS-CONF-2016-045.

\bibitem{CMS:2016abv}
{\bfseries CMS} Collaboration, C.~Collaboration, ``{Search for a high-mass
  resonance decaying into a dilepton final state in 13 fb$^{-1}$ of pp
  collisions at $\sqrt{s}=13~\mathrm{TeV}$},''
CMS-PAS-EXO-16-031.

\bibitem{LEP:2003aa}
{\bfseries SLD Electroweak Group, SLD Heavy Flavor Group, DELPHI, LEP, ALEPH,
  OPAL, LEP Electroweak Working Group, L3} Collaboration, t.~S. Electroweak,
  ``{A Combination of preliminary electroweak measurements and constraints on
  the standard model},''
\href{http://arxiv.org/abs/hep-ex/0312023}{{\ttfamily arXiv:hep-ex/0312023
  [hep-ex]}}.

\bibitem{Carena:2004xs}
M.~Carena, A.~Daleo, B.~A. Dobrescu, and T.~M.~P. Tait, ``{$Z^\prime$ gauge
  bosons at the Tevatron},''
  \href{http://dx.doi.org/10.1103/PhysRevD.70.093009}{Phys. Rev. {\bfseries
  D70} (2004) 093009},
\href{http://arxiv.org/abs/hep-ph/0408098}{{\ttfamily arXiv:hep-ph/0408098
  [hep-ph]}}.

\bibitem{Schael:2013ita}
{\bfseries DELPHI, OPAL, LEP Electroweak, ALEPH, L3} Collaboration, S.~Schael
  {\em et~al.}, ``{Electroweak Measurements in Electron-Positron Collisions at
  W-Boson-Pair Energies at LEP},''
  \href{http://dx.doi.org/10.1016/j.physrep.2013.07.004}{Phys. Rept. {\bfseries
  532} (2013) 119--244},
\href{http://arxiv.org/abs/1302.3415}{{\ttfamily arXiv:1302.3415 [hep-ex]}}.

\bibitem{Buttazzo:2013uya}
D.~Buttazzo, G.~Degrassi, P.~P. Giardino, G.~F. Giudice, F.~Sala, A.~Salvio,
  and A.~Strumia, ``{Investigating the near-criticality of the Higgs boson},''
  \href{http://dx.doi.org/10.1007/JHEP12(2013)089}{JHEP {\bfseries 12} (2013)
  089},
\href{http://arxiv.org/abs/1307.3536}{{\ttfamily arXiv:1307.3536 [hep-ph]}}.

\bibitem{Coriano:2014mpa}
C.~Coriano, L.~Delle~Rose, and C.~Marzo, ``{Vacuum Stability in U(1)-Prime
  Extensions of the Standard Model with TeV Scale Right Handed Neutrinos},''
  \href{http://dx.doi.org/10.1016/j.physletb.2014.09.001}{Phys. Lett.
  {\bfseries B738} (2014) 13--19},
\href{http://arxiv.org/abs/1407.8539}{{\ttfamily arXiv:1407.8539 [hep-ph]}}.

\bibitem{DiChiara:2014wha}
S.~Di~Chiara, V.~Keus, and O.~Lebedev, ``{Stabilizing the Higgs potential with
  a Z$'$},'' \href{http://dx.doi.org/10.1016/j.physletb.2015.03.013}{Phys.
  Lett. {\bfseries B744} (2015) 59--66},
\href{http://arxiv.org/abs/1412.7036}{{\ttfamily arXiv:1412.7036 [hep-ph]}}.

\bibitem{Coriano:2015sea}
C.~Coriano, L.~Delle~Rose, and C.~Marzo, ``{Constraints on abelian extensions
  of the Standard Model from two-loop vacuum stability and $U(1)_{B-L}$},''
  \href{http://dx.doi.org/10.1007/JHEP02(2016)135}{JHEP {\bfseries 02} (2016)
  135},
\href{http://arxiv.org/abs/1510.02379}{{\ttfamily arXiv:1510.02379 [hep-ph]}}.

\bibitem{Casas:2004gh}
J.~A. Casas, J.~R. Espinosa, and I.~Hidalgo, ``{Implications for new physics
  from fine-tuning arguments. 1. Application to SUSY and seesaw cases},''
  \href{http://dx.doi.org/10.1088/1126-6708/2004/11/057}{JHEP {\bfseries 11}
  (2004) 057},
\href{http://arxiv.org/abs/hep-ph/0410298}{{\ttfamily arXiv:hep-ph/0410298
  [hep-ph]}}.

\bibitem{Bell:2016uhg}
N.~F. Bell, Y.~Cai, and R.~K. Leane, ``{Impact of mass generation for spin-1
  mediator simplified models},''
  \href{http://dx.doi.org/10.1088/1475-7516/2017/01/039}{JCAP {\bfseries 1701}
  no.~01, (2017) 039},
\href{http://arxiv.org/abs/1610.03063}{{\ttfamily arXiv:1610.03063 [hep-ph]}}.

\bibitem{Barger:1980ti}
V.~D. Barger, W.-Y. Keung, and E.~Ma, ``{Doubling of Weak Gauge Bosons in an
  Extension of the Standard Model},''
\href{http://dx.doi.org/10.1103/PhysRevLett.44.1169}{Phys. Rev. Lett.
  {\bfseries 44} (1980) 1169}.

\bibitem{Pumplin:2002vw}
J.~Pumplin, D.~R. Stump, J.~Huston, H.~L. Lai, P.~M. Nadolsky, and W.~K. Tung,
  ``{New generation of parton distributions with uncertainties from global QCD
  analysis},'' \href{http://dx.doi.org/10.1088/1126-6708/2002/07/012}{JHEP
  {\bfseries 07} (2002) 012},
\href{http://arxiv.org/abs/hep-ph/0201195}{{\ttfamily arXiv:hep-ph/0201195
  [hep-ph]}}.

\bibitem{Aprile:2017iyp}
{\bfseries XENON} Collaboration, E.~Aprile {\em et~al.}, ``{First Dark Matter
  Search Results from the XENON1T Experiment},''
\href{http://arxiv.org/abs/1705.06655}{{\ttfamily arXiv:1705.06655
  [astro-ph.CO]}}.

\bibitem{Akerib:2016vxi}
{\bfseries LUX} Collaboration, D.~S. Akerib {\em et~al.}, ``{Results from a
  search for dark matter in the complete LUX exposure},''
  \href{http://dx.doi.org/10.1103/PhysRevLett.118.021303}{Phys. Rev. Lett.
  {\bfseries 118} no.~2, (2017) 021303},
\href{http://arxiv.org/abs/1608.07648}{{\ttfamily arXiv:1608.07648
  [astro-ph.CO]}}.

\bibitem{Szydagis:2016few}
{\bfseries LUX, LZ} Collaboration, M.~Szydagis, ``{The Present and Future of
  Searching for Dark Matter with LUX and LZ},'' PoS {\bfseries ICHEP2016}
  (2016) 220,
\href{http://arxiv.org/abs/1611.05525}{{\ttfamily arXiv:1611.05525
  [astro-ph.CO]}}.

\bibitem{Charles:2016pgz}
{\bfseries Fermi-LAT} Collaboration, E.~Charles {\em et~al.}, ``{Sensitivity
  Projections for Dark Matter Searches with the Fermi Large Area Telescope},''
  \href{http://dx.doi.org/10.1016/j.physrep.2016.05.001}{Phys. Rept. {\bfseries
  636} (2016) 1--46},
\href{http://arxiv.org/abs/1605.02016}{{\ttfamily arXiv:1605.02016
  [astro-ph.HE]}}.

\bibitem{daSilva:2017swg}
{\bfseries LUX} Collaboration, C.~F.~P. da~Silva, ``{Dark Matter Searches with
  LUX},''
\newblock 2017.
\newblock
\href{http://arxiv.org/abs/1710.03572}{{\ttfamily arXiv:1710.03572 [hep-ex]}}.
\newblock

\bibitem{Aartsen:2016zhm}
{\bfseries IceCube} Collaboration, M.~G. Aartsen {\em et~al.}, ``{Search for
  annihilating dark matter in the Sun with 3 years of IceCube data},''
  \href{http://dx.doi.org/10.1140/epjc/s10052-017-4689-9}{Eur. Phys. J.
  {\bfseries C77} no.~3, (2017) 146},
\href{http://arxiv.org/abs/1612.05949}{{\ttfamily arXiv:1612.05949
  [astro-ph.HE]}}.

\bibitem{Ohki:2008ff}
H.~Ohki, H.~Fukaya, S.~Hashimoto, T.~Kaneko, H.~Matsufuru, J.~Noaki, T.~Onogi,
  E.~Shintani, and N.~Yamada, ``{Nucleon sigma term and strange quark content
  from lattice QCD with exact chiral symmetry},''
  \href{http://dx.doi.org/10.1103/PhysRevD.78.054502}{Phys. Rev. {\bfseries
  D78} (2008) 054502},
\href{http://arxiv.org/abs/0806.4744}{{\ttfamily arXiv:0806.4744 [hep-lat]}}.

\bibitem{Crewther:1972kn}
R.~J. Crewther, ``{Nonperturbative evaluation of the anomalies in low-energy
  theorems},''
\href{http://dx.doi.org/10.1103/PhysRevLett.28.1421}{Phys. Rev. Lett.
  {\bfseries 28} (1972) 1421}.

\bibitem{Chanowitz:1972vd}
M.~S. Chanowitz and J.~R. Ellis, ``{Canonical Anomalies and Broken Scale
  Invariance},''
\href{http://dx.doi.org/10.1016/0370-2693(72)90829-5}{Phys. Lett. {\bfseries
  40B} (1972) 397--400}.

\bibitem{Chanowitz:1972da}
M.~S. Chanowitz and J.~R. Ellis, ``{Canonical Trace Anomalies},''
\href{http://dx.doi.org/10.1103/PhysRevD.7.2490}{Phys. Rev. {\bfseries D7}
  (1973) 2490--2506}.

\bibitem{Collins:1976yq}
J.~C. Collins, A.~Duncan, and S.~D. Joglekar, ``{Trace and Dilatation Anomalies
  in Gauge Theories},''
\href{http://dx.doi.org/10.1103/PhysRevD.16.438}{Phys. Rev. {\bfseries D16}
  (1977) 438--449}.

\bibitem{Shifman:1978zn}
M.~A. Shifman, A.~I. Vainshtein, and V.~I. Zakharov, ``{Remarks on Higgs Boson
  Interactions with Nucleons},''
\href{http://dx.doi.org/10.1016/0370-2693(78)90481-1}{Phys. Lett. {\bfseries
  78B} (1978) 443--446}.

\end{thebibliography}\endgroup


\end{document}